\documentclass[twocolumn,pra,aps,groupedaddress,showpacs,amsmath,amssymb,10pt,longbibliography]{revtex4-2}
\usepackage{amsmath} 
\usepackage{amssymb} 
\usepackage{amsfonts} 
\usepackage{bm} 
\usepackage{graphicx} 
\usepackage{dcolumn} 
\usepackage{newtxtext}
\usepackage{newtxmath}
\usepackage{wasysym} 
\usepackage{makeidx}
\usepackage{color}
\usepackage{mathtools}
\usepackage{threeparttable}
\usepackage[linkcolor=blue,anchorcolor=black,citecolor=blue,urlcolor=blue,colorlinks=true]{hyperref}
\usepackage{breakurl}
\usepackage{float}
\usepackage{soul}
\usepackage{adjustbox}
\usepackage{autobreak}  

\begin{document}

\title{Optical pumping of alkali-metal vapor in the quasi-high-pressure regime}

\author{Kezheng Yan}
\author{Jinbo Hu}
\author{Nan Zhao}
\email{nzhao@csrc.ac.cn}
\affiliation{
    \href{https://ror.org/04tavf782}{Beijing Computational Science Research Center}, 
    Beijing 100193, People's Republic of China
}
\date{\today}

\begin{abstract}
Optical pumping is fundamental to high-precision measurement using thermal alkali-metal atoms in vapor cells. 
In applications such as atomic magnetometry, buffer gases (e.g., $\mathrm{N}_2$ or $\mathrm{He}$) at specific pressures are introduced to quench fluorescence and mitigate wall relaxation. 
In the high-pressure limit (e.g., the $\mathrm{N}_2$ pressure $p_{\mathrm{N}_2}> 1$~atm), where collisional broadening exceeds hyperfine splittings of the atoms, 
optical pumping theory provides a clear description of the angular momentum exchange between photons and atomic spins. 
However, in many magnetic sensing scenarios, the high-pressure approximation becomes inadequate as its pressure conditions are not strictly satisfied. 
Consequently, an explicit description of optical pumping under realistic pressures is critical for selecting operating points and enhancing system performance. 
To address this, we develop a unified theoretical framework of optical pumping in the quasi-high-pressure regime, 
where collisional broadening is comparable to the ground-state hyperfine splitting. 
We demonstrate that light absorption, spin polarization, and magnetic-resonance linewidth in this regime differ significantly from those predicted by the high-pressure limit and offer favorable operating conditions. 
Our study extends conventional modeling and offers critical guidance for atomic magnetometry operating under realistic buffer gas pressures.
\end{abstract}

\maketitle

\section{\label{sec:introduction}Introduction}
Optical pumping of alkali-metal atoms serves as the foundation for a wide array of quantum sensing technologies, ranging from atomic clocks \cite{JauClock2004, VanierBook1989} to magnetometers \cite{BudkerOM2007,BudkerBook2013,FangSensors2012} and inertial sensors \cite{WrightInertial2022,WalkerGyroBook2016,Gao2024}. In particular, atomic magnetometers exploit the spin dynamics of alkali-metal atoms to achieve ultrasensitive field detection. In this process, a circularly polarized light beam creates atomic spin polarization, which then precesses at the Larmor frequency in an external magnetic field, providing a precise measurement of the field strength.

Vapor cells used in these applications are generally categorized as either evacuated or buffer-gas-filled \cite{HapperRMP1972}. While evacuated cells rely on anti-relaxation coatings to prevent wall-collision relaxation \cite{Seltzer2008,SeltzerCoating2009}, buffer-gas-filled cells utilize inert gases to suppress atomic diffusion and quench fluorescence \cite{WagshulLayer1994,LancorPressure2010}. These features allow operation at higher temperatures and atomic densities, enabling the ultra-high sensitivity characteristic of modern magnetometers \cite{XiaoPRL2024,LuSERFBufferGas2022,JiaIEEE2021,SchoenauVector2025}. This work focuses specifically on the optical pumping dynamics within such buffer-gas-filled cells.

Depending on the buffer gas pressure, optical pumping operates in distinct regimes. 
In the low-pressure (LP) regime with typical buffer gas pressure $\lesssim 10~\mathrm{Torr}$, 
strong and narrow optical resonances of well resolved hyperfine multiplets enable refined optical manipulation, such as in atomic clocks based on coherent population trapping (CPT) \cite{VanierCPT2005}. 
In the high-pressure (HP) limit (e.g., buffer gas pressure $> 1~\mathrm{atm}$), strong collisional quenching and pressure broadening dominate \cite{SavukovRF2005,HapperBook2010,WalkerSEOP1997,Tang2025}. This treatment renders hyperfine multiplets unresolved and
forms the basis of the spin-temperature distribution (STD) widely used in, e.g., spin-exchange-relaxation-free (SERF) magnetometers \cite{HapperSERF1977,SavukovSERF2005}. 
However, many optical pumping systems including atomic magnetometers operate in the quasi-high-pressure (QHP) regime (typically $\sim 10^2~\mathrm{Torr}$) \cite{ChangPRA2019,ScholtesMx2011,SchultzeLSDMz2017,JiangPhotonics2026}.
We focus on this intermediate regime, in which the ground-state (GS) hyperfine structure remains spectrally resolvable while the excited-state (ES) structure is unresolved. 
This leads to spin dynamics that are neither fully captured by the LP models nor the simplified HP approximations.
Developing a precise quantitative understanding of the QHP regime is therefore critical. It not only enables the determination of optimal operating points to maximize sensitivity, but also facilitates the assessment of long-term system stability against parameter drifts.

The spectral resolvability of the GS hyperfine structure fundamentally governs the optical pumping process in the QHP regime. Unlike the HP limit, the QHP regime requires a rigorous treatment of the resolvable multiplets and their distinct contributions to absorption. Starting from the master equation in Liouville space, we incorporate the electric-dipole interaction and relevant collisional mechanisms to derive the analytical formalism of optical pumping in the QHP regime. Crucially, we establish a compact relation linking QHP superoperators to their HP counterparts, providing a unified theoretical framework that facilitates both physical analysis and efficient numerical evaluation.

A distinct feature of the QHP regime is the hyperfine-resolved dependence of absorption on pumping-light polarization and intensity, which markedly differs from the HP limit. Within our framework, we derive analytical expressions for absorption cross section under both linearly and circularly polarized pumping. We elucidate how the redistribution of population among and within hyperfine multiplets reshapes the detuning-dependent absorption profile.

Consequently, driven by the combined redistribution dynamics in the QHP regime, the steady-state atomic population generally deviates from the standard STD. We demonstrate that while the STD can approximate the distribution shape under rapid spin-exchange collisions, the effective temperature parameter—--namely the spin polarization—--requires QHP-specific corrections. Evaluating these dynamics, our results indicate that resonant pumping at specific hyperfine level can yield a larger polarization.

Beyond these fundamental spin dynamics, we analyze the magnetic-resonance linewidth to identify optimal operating conditions for atomic magnetometers. We find that resonant pumping at the lower hyperfine multiplet not only enhances spin polarization but also yields a narrower magnetic-resonance linewidth. This suggests a strategy for optimizing sensitivity through simultaneous improvements in resonance amplitude and linewidth.

The remainder of this paper is organized as follows. Section~\ref{sec:pressure_regimes} defines the regimes of buffer gas pressure based on spectral characteristics. Section~\ref{sec:theoretical_treatment} presents the QHP-specific theoretical framework based on the Liouville space master equation. Section~\ref{sec:results_and_discussion} discusses the unique features of absorption, spin polarization, and magnetic-resonance linewidth in the QHP regime. Finally, Section~\ref{sec:conclusions} summarizes the main findings.

\section{\label{sec:pressure_regimes}Regimes of buffer gas pressure}

\begin{figure} 
\centering 
\includegraphics[width=1\linewidth]{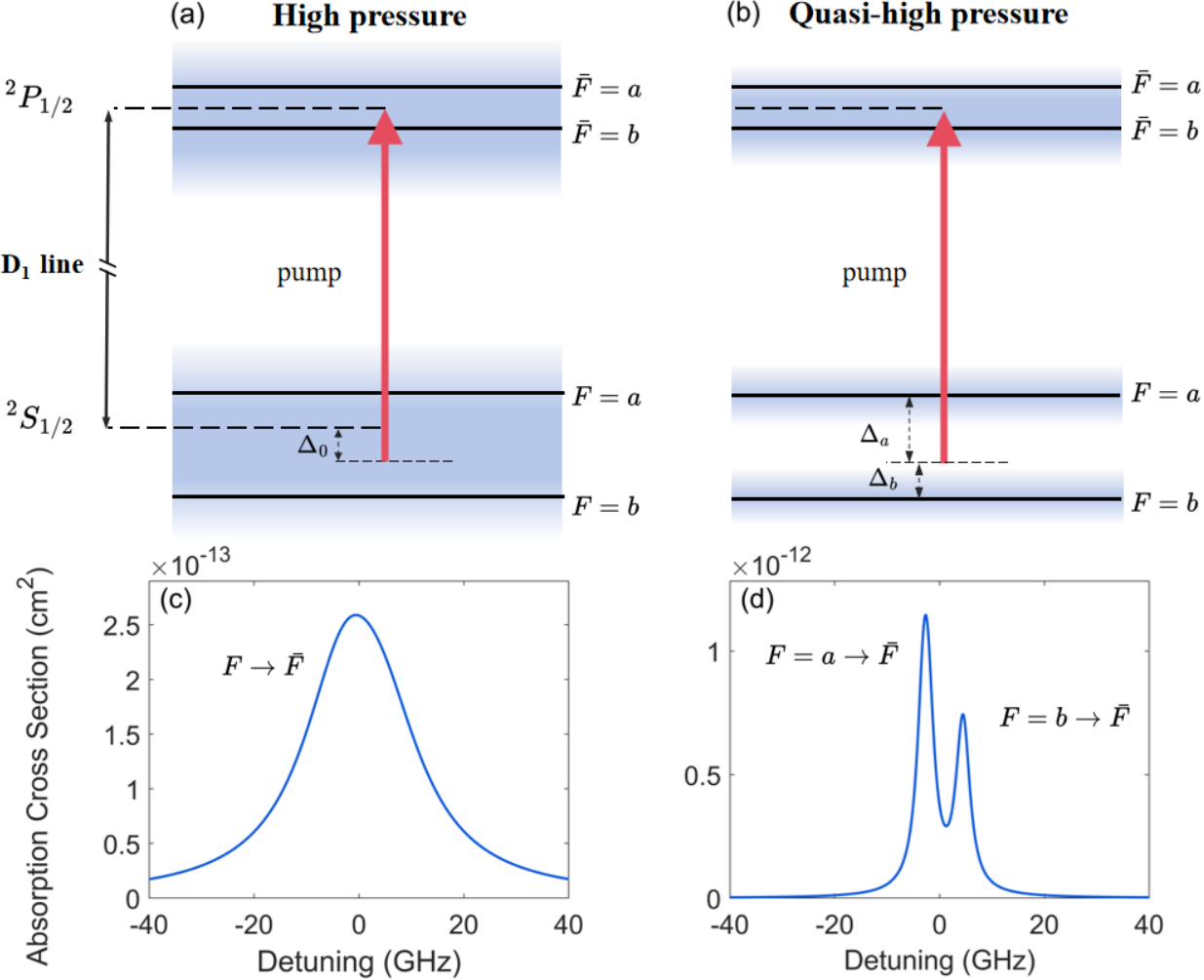} 
\caption{
Schematic illustration of optical pumping on the alkali-metal D$_1$ line in the HP limit and QHP regime. 
(a) and (b) show the GS and ES hyperfine multiplets, with different collisional broadening $\Gamma_{\mathrm{brd}}$ denoted by blue level edges. Each of $F$ and $\bar{F}$ corresponds to either $a=I+1/2$ or $b=I-1/2$ for nuclear spin $I$. Light detuning from the D$_1$ line reference frequency is denoted by $\Delta_0$, while $\Delta_{F}$ for $F=a$ and $b$ are the detunings shifted from $\Delta_0$ by the GS hyperfine splitting.  
(c) and (d) present the corresponding absorption cross section for unpolarized $^{87}$Rb, with $2\Gamma_{\mathrm{brd}}=20$ and $3~\mathrm{GHz}$, respectively.
Doppler broadening $\Gamma_{\mathrm{dop}} = 0.57~\mathrm{GHz}$ is included using the Voigt profile, which is negligible in both regimes. 
Here, collisional shifts are not included for simplicity.
}
\label{fig1:pumping_schematic} 
\end{figure}

Optical pumping of alkali-metal atoms relies on the coupling between the GS and ES of the atomic D line transition, which drives atoms toward a targeted state via optical excitation and collisional relaxation.
Taking the $\mathrm{D}_1$ transition as an example, Fig.~\ref{fig1:pumping_schematic} shows that while light absorption initiates the pumping process, the spectral characteristics are governed by line broadening mechanisms.
In buffer-gas vapor cells, which are the focus of this work, collisional broadening $\Gamma_{\mathrm{brd}}$ dominates, serving as the primary criterion for distinguishing pressure regimes.

To characterize these regimes, we analyze the absorption cross section for unpolarized atoms in the limit of vanishing photon flux $\Phi$ of the incident laser beam (i.e., $\Phi \rightarrow 0$). 
In the LP regime, $\Gamma_{\mathrm{brd}}$ is insufficient to bridge the ES or GS hyperfine splitting, and dynamics are further complicated by radiation trapping \cite{LancorPressure2010} and atomic diffusion \cite{IshikawaDiffusion2000,ChangPRA2024}. 
In the opposite HP limit,  $\Gamma_{\mathrm{brd}}$ is sufficiently large to merge all GS and ES hyperfine splittings into a single resonance peak, as depicted in Figs.~\ref{fig1:pumping_schematic}(a) and \ref{fig1:pumping_schematic}(c).
In this case, the system simplifies significantly, and underlies the STD formalism \cite{HapperSERF1977,Appelt1998}.

The QHP regime represents an intermediate condition where $\Gamma_{\mathrm{brd}}$ is comparable to the GS hyperfine splitting but significantly larger than that of the ES.
Figures~\ref{fig1:pumping_schematic}(b) and \ref{fig1:pumping_schematic}(d) illustrate that the absorption profile simplifies to two resolved components associated with the GS multiplets $F=a$ and $b$.
Unlike the HP limit, where atoms in different multiplets absorb photons at an almost common rate, in the QHP regime, the pumping process becomes more sensitive to the pump laser frequency.
Furthermore, stronger pump light significantly alters the population distribution between the $F=a$ and $b$ levels, which in turn reshapes the frequency-selective absorption behavior and spin polarization—--factors critical for subsequent applications.

While the linear absorption profile in the $\Phi \rightarrow 0$ limit serves as a convenient classification metric, the rich physical phenomena emerging at higher optical intensities are the primary focus of this work.
We emphasize that the simple optical pumping model used in the HP limit is inadequate to describe the behaviors in the QHP regime.
The following sections will present a unified theoretical framework capable of treating the QHP regime,
and analyze the spin evolution in practical scenarios, such as atomic magnetometers operating in intermediate pressure.

\section{\label{sec:theoretical_treatment}THEORETICAL TREATMENT}

\subsection{\label{sec:master_equation}Master equation in Liouville space}

We study the spin dynamics of optically pumped alkali-metal atoms in vapor cells, where both coherent and incoherent processes contribute to the evolution of the atomic ensemble.
For the coherent part, the atomic spin Hamiltonian is \cite{HapperRMP1972}
\begin{align}
\label{eq:Hamiltonian_0}
H_0 &=\frac{\omega^{\{\mathrm{eg}\}}}{2}\left(1^{\{\mathrm{e}\}} - 1^{\{\mathrm{g}\}}\right) + 
H^{\{\mathrm{g}\}} + H^{\{\mathrm{e}\}}, \\
\label{eq:Hamiltonian_g}
H^{\{\mathrm{g}\}} &= \sum_{F,m_F} \omega^{\{\mathrm{g}\}}_{F,m_F}|F,m_F\rangle\langle F,m_F|, \\
\label{eq:Hamiltonian_e}
H^{\{\mathrm{e}\}} &= \sum_{\bar{F}, \bar{m}_F} \omega^{\{\mathrm{e}\}}_{\bar{F},\bar{m}_{F}}|\bar{F},\bar{m}_F\rangle\langle \bar{F},\bar{m}_F|, 
\end{align}
where $1^{\{\mathrm{g}\}}$ and $1^{\{\mathrm{e}\}}$ are identity operators of the GS and ES subspaces, respectively, and $\omega^{\{\mathrm{eg}\}}$ denotes the reference frequency of the optical transition.
In Eqs.~\eqref{eq:Hamiltonian_g} and \eqref{eq:Hamiltonian_e}, the Hamiltonians $H^{\{\mathrm{g}\}}$ and $H^{\{\mathrm{e}\}}$ are expressed in their eigen-representations with eigen-frequencies $\omega^{\{\mathrm{g}\}}_{F,m_F}$ and $\omega^{\{\mathrm{e}\}}_{\bar{F},\bar{m}_{F}}$, together with the eigen-states $|F,m_F\rangle$ and $|\bar F,\bar m_F\rangle$, respectively.
In this paper, we assume the magnetic field is so weak that the total angular momenta $F$ and $\bar{F}$ are approximately good quantum numbers, 
with corresponding magnetic quantum numbers $m_F$ and $\bar{m}_F$. 

Having specified the atomic Hamiltonian $H_0$, we introduce the light-atom interaction Hamiltonian $H_{\mathrm{int}}$, which takes an electric-dipole form \cite{HapperRMP1972}
\begin{align}
H_{\mathrm{int}}
= V+V^{\dagger}
= -\Omega_{\mathrm{I}}\mathbf{d}^{\dagger}\cdot\hat{\mathbf{e}} -\Omega^*_{\mathrm{I}}\mathbf{d}\cdot\hat{\mathbf{e}}^{*}, 
\label{eq:H_int}
\end{align}
with $V$ describing photon absorption and atom excitation with optical Rabi frequency $\Omega_{\mathrm{I}}$, and $V^{\dagger}$ representing the reverse process. 
Here, $\mathbf{d}$ denotes the dimensionless dipole operators, 
and $\hat{\mathbf{e}}$ is the unit vector of the positive-frequency classical electric field.
For a monochromatic laser beam with frequency $\omega$, the laser detuning with respect to the atomic transition $\vert F, m_F\rangle \leftrightarrow \vert \bar{F}, \bar{m}_F\rangle$ is
\begin{align}
\Delta_{\bar{F}\bar{m}_F,Fm_{F}} &= \Delta_0 - \left(\omega^{\{\mathrm{e}\}}_{\bar{F},\bar{m}_{F}} - \omega^{\{\mathrm{g}\}}_{F,m_F} \right),
\label{eq:Delta}
\end{align}
where $\Delta_0 = \omega - \omega^{\{\mathrm{eg}\}}$ is the hyperfine-splitting-independent detuning of the laser frequency $\omega$ with respect to the reference frequency of the D line transition $\omega^{\{\mathrm{eg}\}}$.

In addition to the coherent mechanisms governed by the Hamiltonians $H_0$ and $H_{\mathrm{int}}$, 
the atoms experience incoherent dynamics, including level broadening and spin relaxation due to atomic collisions, spontaneous emission of photons, and non-radiative decay (quenching) processes. 
Due to these relaxation processes, the optical ES are hardly occupied. 
In this case, the ES degrees of freedom and the optical coherences are adiabatically eliminated \cite{HapperRMP1972}.
Therefore, the evolution of atomic spin is reduced to the master equation in GS Liouville space \cite{Appelt1998}
\begin{align}
\frac{d}{dt}\left|{\rho}\right) = 
&-\left(iH^{\{\mathrm{g}\}\copyright}+
\mathcal{G}^{\{\mathrm{g}\}}_{\mathrm{rel}} + 
R_{\mathrm{op}}\mathcal{A}_{\mathrm{op}} \right)\left|\rho\right), 
\label{eq:effective_masterEq}
\end{align}
where $\left|\rho\right)$ is the vectorized GS density matrix, and the notation $X^{\copyright} = X^{\flat} - X^{\dagger\sharp}$ is the commutator superoperator
with $X^{\flat}$ and $X^{\sharp}$ being the left- and right-translation superoperators \cite{HapperBook2010}, 
corresponding to the operator $X$ in Hilbert space.  
The GS spin relaxation term $\mathcal{G}^{\{\mathrm{g}\}}_{\mathrm{rel}}$ is
\begin{align}
\mathcal{G}^{\{\mathrm{g}\}}_{\mathrm{rel}} &= 
\Gamma_{\mathrm{SD}} \mathcal{A}_{\mathrm{SD}} + \Gamma_{\mathrm{EX}} \left(\mathcal{A}_{\mathrm{SD}} - 
(\mathbf{S}|{\rho})\cdot  \boldsymbol{\mathcal{A}}_{\mathrm{SE}} \right), 
\end{align}
with the spin-destruction rate $\Gamma_{\mathrm{SD}}$, the spin-exchange rate $\Gamma_{\mathrm{EX}}$, 
and the expectation value of the electron spin vector $(\mathbf{S} \vert \rho) = \mathrm{Tr}\left[\mathbf{S}{\rho}\right]$. 
The spin-destruction and spin-exchange superoperators $\mathcal{A}_{\mathrm{SD}}$ and $\mathcal{A}_{\mathrm{SE}}$ are
\begin{align}
\label{eq:SD}
\mathcal{A}_{\mathrm{SD}} &=  \frac{3}{4} - {\mathbf{S}}^{\flat} \cdot {\mathbf{S}}^{\sharp} ,\\
\label{eq:A_SE}
{\mathcal{A}}_{\mathrm{SE}} &= 
{\mathbf{S}}^{\flat} + {\mathbf{S}}^{\sharp} 
-2i {\mathbf{S}}^{\flat} \times {\mathbf{S}}^{\sharp}, 
\end{align}
where $\mathbf{S}$ is the electron spin operator of the alkali-metal atom. Physically, $\mathcal{A}_{\mathrm{SD}}$ relaxes the mean electron spin $(\mathbf{S}\vert{}\rho)$ toward complete depolarization, whereas $\mathcal{A}_{\mathrm{SE}}$ originates from the exchange of electron spins during collisions, a process through which $(\mathbf{S}\vert{}\rho)$ is conserved.

The optical pumping process is described by the third term of the right-hand-side of Eq.~\eqref{eq:effective_masterEq}, where $R_{\mathrm{op}}$ is the optical pumping rate and $\mathcal{A}_{\mathrm{op}}$ is the corresponding superoperator.
This superoperator describes the overall influence of the pumping light on the GS density matrix, including light absorption, light shift, population redistribution, and evolution of coherences.
The optical pumping process consists of two parts \cite{HapperRMP1972, HapperBook2010}
\begin{align}
\mathcal{A}_{\mathrm{op}} = \mathcal{A}_{\mathrm{dp}} - \mathcal{A}_{\mathrm{rp}},
\label{eq:op_dp+rp}
\end{align}
where $\mathcal{A}_{\mathrm{dp}}$ and $\mathcal{A}_{\mathrm{rp}}$ represent the depopulation and repopulation processes, respectively.
The depopulation term $\mathcal{A}_{\mathrm{dp}}$ is associated with light absorption and light shift
\begin{align}
R_{\mathrm{op}}\mathcal{A}_{\mathrm{dp}}&= 
i[ V^{\dagger}W ]^{\copyright}\equiv i[\delta H^{\{\mathrm{g}\}}]^{\copyright}, 
\label{eq:dp_def}
\end{align}
where 
\begin{align}
\label{eq:W}
W &=V./(\Delta^{\{\mathrm{eg}\}}+i\Gamma_{\mathrm{brd}}).
\end{align}
In Eq.~\eqref{eq:W}, $\Delta^{\{\mathrm{eg}\}} = \left[ \Delta_{\bar{F}\bar{m}_F, Fm_F}\right]$ is the detuning matrix with elements defined in Eq.~\eqref{eq:Delta}, 
and the dot–slash product is defined by 
$\langle\nu \vert A./B \vert \mu\rangle = \langle\nu \vert A \vert \mu\rangle/\langle\nu \vert B \vert \mu\rangle$
for matrices $A$ and $B$.
The repopulation term $\mathcal{A}_{\mathrm{rp}}$, which includes optical excitation, spin relaxation within the ES and quenching transition, takes the following form \cite{Appelt1998, HapperBook2010}
\begin{align}
\mathcal{A}_{\mathrm{rp}} = & \left(\Gamma_{\mathrm{s}} \mathcal{A}_{\mathrm{s}}^{\{\mathrm{ge}\}}+ \Gamma_{\mathrm{q}} \mathcal{A}_{\mathrm{q}}^{\{\mathrm{ge}\}} \right) \nonumber \\
& \times \left({iH^{\{\mathrm{e}\}\copyright}} + \Gamma_{\mathrm{s}} + \Gamma_{\mathrm{q}}+\Gamma_{\mathrm{JD}} \mathcal{A}_{\mathrm{JD}}\right)^{-1} \mathcal{A}_{\mathrm{p}}^{\{\mathrm{eg}\}},
\label{eq:rp_def}
\end{align}
where $\mathcal{A}_{\mathrm{s}}^{\{\mathrm{ge}\}}$ and $\mathcal{A}_{\mathrm{q}}^{\{\mathrm{ge}\}}$ are the superoperators of the spontaneous emission and quenching processes with the corresponding rates $\Gamma_{\mathrm{s}}$ and $\Gamma_{\mathrm{q}}$.
The J-destruction superoperator $\mathcal{A}_{\mathrm{JD}}$ describes the relaxation of the mean ES angular momentum $\langle \mathbf{J} \rangle$ due to spin-destruction collisions, analogous to the GS $\mathcal{A}_{\mathrm{SD}}$ but at a much faster rate $\Gamma_{\mathrm{JD}}$ \cite{HapperBook2010}.
The analytical expressions of $\mathcal{A}_{\mathrm{s}}^{\{\mathrm{ge}\}}$ , $\mathcal{A}_{\mathrm{q}}^{\{\mathrm{ge}\}}$ and $\mathcal{A}_{\mathrm{JD}}$ are derived in Refs.~\cite{Appelt1998, HapperBook2010, LancorPressure2010}. 
The superoperator $\mathcal{A}_{\mathrm{p}}^{\{\mathrm{eg}\}}$ in Eq.~\eqref{eq:rp_def} is defined as
\begin{align}
R_{\mathrm{op}}\mathcal{A}_{\mathrm{p}}^{\{\mathrm{eg}\}} &= 
i\left(V^{*} {\otimes} W 
-W^{*} \otimes V\right),
\label{eq:Ap_eg_def}
\end{align}
which represents the optical excitation of the atoms from the GS to the ES. 
Here, $\otimes$ denotes the Kronecker product between operators in Hilbert space.

\subsection{\label{sec:QHP_pumping}Optical pumping in quasi-high-pressure regime}

\subsubsection{\label{sec:depopulation}Depopulation}
The energy denominator matrices
$\Delta^{\{\mathrm{eg}\}} + i\Gamma_{\mathrm{brd}}$ in Eq.~\eqref{eq:W} play a central role in describing optical pumping in different buffer-gas regimes.  
In the HP limit, the collisional broadening $\Gamma_{\mathrm{brd}}$ is much larger than the difference between the elements of the detuning matrix $\Delta^{\{\mathrm{eg}\}}$, i.e., $\Gamma_{\mathrm{brd}}\gg \left\vert \omega_{\bar{F},\bar{m}_F}^{\{\mathrm{e}\}} - \omega_{F, m_F}^{\{\mathrm{g}\}}\right\vert$. In this case, $\Delta^{\{\mathrm{eg}\}}$ is simplified to a single-valued matrix
\begin{align}
\Delta^{\{\mathrm{eg}\}} \approx \Delta_0 \sum_{\bar{F}, \bar{m}_F, F, m_F} \vert \bar{F},\bar{m}_F\rangle \langle F, m_F\vert,
\label{eq:highP_approx}
\end{align}
and the effective Hamiltonian in Eq.~\eqref{eq:dp_def} becomes
\begin{align}
\delta H^{\{\mathrm{g}\}}= \frac{V^{\dagger}V}{\Delta_0 + i\Gamma_{\mathrm{brd}}}.
\label{eq:highP_approx2}
\end{align}
Physically, the approximation in Eq.~\eqref{eq:highP_approx} means that, in the large broadening limit, 
the different detuning frequencies have no discriminatory effect on sublevel distinction within the pumping process. 
Consequently, the alkali-metal atom exhibits a two-level-like spectral response. 
With the symmetry property of the electric dipole operator $\hat{\mathbf{e}}^* \cdot \mathbf{d} \mathbf{d}^{\dagger} \cdot \hat{\mathbf{e}}= 1/2 - K\mathbf{s} \cdot \mathbf{S}$,
the HP approximation Eqs.~\eqref{eq:highP_approx} and \eqref{eq:highP_approx2} yields a compact form of the depopulation pumping superoperator \cite{HapperBook2010}, 
\begin{align}
\label{eq:A_dp_HP}
\mathcal{A}^{(\mathrm{HP})}_{\mathrm{dp}}&=  1-K\mathbf{s}\cdot \left(\mathbf{S}^{\flat}+\mathbf{S}^{\sharp} \right)  -\frac{i\Delta_0}{\Gamma_{\mathrm{brd}}}  K\mathbf{s}\cdot \mathbf{S}^{
\copyright},
\end{align}
where $K=1$ corresponds to D$_1$-line pumping, with $K=-1/2$ for the D$_2$ line, and $\mathbf{s}=i\hat{\mathbf{e}}\times\hat{\mathbf{e}}^{*}$ is the effective photon spin of the pumping light. 
The real and imaginary parts of $\mathcal{A}_{\mathrm{dp}}^{(\mathrm{HP})}$ describe light absorption and optical shift, respectively. 
The overall strength of this depopulation process is quantified by the optical pumping rate $R_{\mathrm{op}} = \bar{\sigma} \Phi$, which is proportional to the incident photon flux $\Phi$, with $\bar{\sigma}$ being the {\it intrinsic} absorption cross section of unpolarized atoms (i.e., all the GS sublevels are equally populated). 
The intrinsic absorption cross section in the HP limit is a function of detuning $\Delta_0$
\begin{align}
\bar{\sigma}_{\mathrm{HP}}  = \frac{A}{\pi} \frac{\Gamma_{\mathrm{brd}}}{\Delta_0^2 + \Gamma_{\mathrm{brd}}^2} \equiv A \mathcal{L}(\Delta_0),
\end{align}
where $\mathcal{L}(\Delta)$ is the normalized Lorentzian line shape function, 
and $A = \pi r_{\mathrm{e}}c f^{\{\mathrm{ge}\}}$ is the area constant with the classical electron radius $r_{\mathrm{e}}$, 
the speed of light $c$ and the oscillator strength of the atom $f^{\{\mathrm{ge}\}}$ \cite{HapperRMP1972}.

\begin{figure}[t!]
\centering
\includegraphics[width=0.95\linewidth]{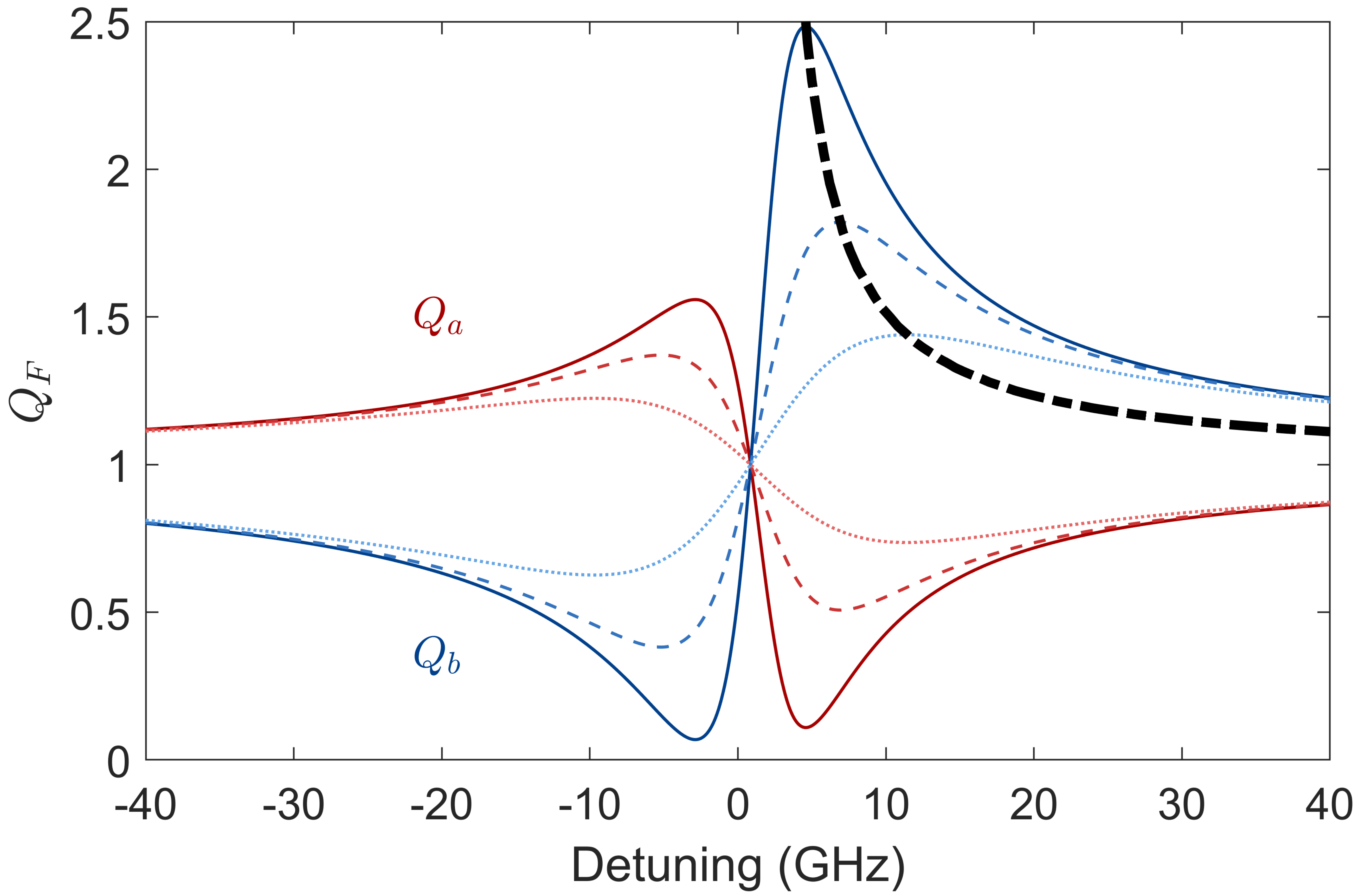}
\caption{
Weighting factor $Q_{F}$ for $F=a$ and $F=b$ [see Eq.~\eqref{eq:Q_F}],
with the GS hyperfine splitting $6.8~\mathrm{GHz}$ and different values of collisional broadening $2\Gamma_{\mathrm{brd}}=3$, $10$, and $20~\mathrm{GHz}$ (solid, dashed, and dotted curves, respectively).
The curves converge to a common intersection at $\Delta_a=\Delta_b$.
The thick dash-dot line shows the maximum attainable $Q_{b}$ as $\Gamma_{\mathrm{brd}}$ increases, recovering the HP limit ($Q_F\to 1$).
}
\label{fig2:Q_F}
\end{figure}

In the QHP regime of interest, the ES hyperfine splittings and all the Larmor frequencies in weak magnetic fields are neglected with respect to the collisional broadening, i.e., $\Gamma_{\mathrm{brd}}\gg \left\vert \omega_{\bar{F},\bar{m}_F}^{\{\mathrm{e}\}} - \omega_{\bar{F}',\bar{m}'_F}^{\{\mathrm{e}\}}\right\vert$,
but the GS hyperfine resolvability is retained.
In this case, the elements of the detuning matrix $\Delta^{\{\mathrm{eg}\}}$ are grouped by two values $\Delta_{F}$ for $F=a$ or $F=b$
\begin{align}
\Delta^{\{\mathrm{eg}\}} \approx \sum_{\bar{F}, \bar{m}_F,F, m_F}  \Delta_F \vert \bar{F},\bar{m}_F \rangle \langle F, m_F\vert,
\label{eq:QHP_approximation}
\end{align}
where $\Delta_{F} = \Delta_0  + \omega^{\{\mathrm{g}\}}_{F,0}$ is the frequency detuning with respect to the GS hyperfine levels. 
With this coarse-grained modeling of the detuning matrix, the effective Hamiltonian $\delta H^{\{\mathrm{g}\}}$ defined in Eq.~\eqref{eq:dp_def} becomes
\begin{align}
\label{eq:delta_H_QHP}
\delta H^{\{\mathrm{g}\}} &= \sum_{F=a, b} \delta H^{\{\mathrm{g}\}}_F \Pi_F,\\
\label{eq:delta_H_F}
\delta H^{\{\mathrm{g}\}}_F &= \frac{V^{\dagger}V }{\Delta_F + i\Gamma_{\mathrm{brd}}},
\end{align}
where $\Pi_F = \sum_{m_F} \vert F, m_F\rangle \langle F, m_F\vert $ is the projection operator of the GS subspace with a given total angular momentum quantum number $F$,
and $\delta H_F^{\{\mathrm{g}\}}$ is the $F$-specific counterpart of the effective Hamiltonian.

With the QHP approximation in Eq.~\eqref{eq:QHP_approximation}, we derive that the depopulation superoperator $\mathcal{A}_{\mathrm{dp}}$ becomes
\begin{align}
\mathcal{A}_{\mathrm{dp}} \mathcal{P}_{\mathrm{Z}} = \sum_F  \left[1- K\mathbf{s}\cdot (\mathbf{S}^{\flat} + \mathbf{S}^{\sharp}) - \frac{i\Delta_F}{\Gamma_{\mathrm{brd}}}K\mathbf{s}\cdot \mathbf{S}^{\copyright}\right]Q_F \mathcal{P}_F.
\label{eq:A_dp_QHP}
\end{align}
See Appendix~\ref{app:simplification} for detailed proof.
In comparison to the HP case Eq.~\eqref{eq:A_dp_HP}, the GS hyperfine resolvability brings about the following corrections:
(i) the overall detuning $\Delta_0$ is replaced by the $F$-resolved detuning $\Delta_F$; 
(ii) a weighting factor $Q_F$ accounting for the $F$-resolved absorption is introduced as
\begin{align}
Q_F(\Delta_F) = \frac{\mathcal{L}(\Delta_F)}{x_a\mathcal{L}(\Delta_a) + x_b\mathcal{L}(\Delta_b)},
\label{eq:Q_F}
\end{align}
where $x_F = (2F+1)/[2(2I+1)]$ is the degeneracy weighting factor;
and (iii) the expression in Eq.~\eqref{eq:A_dp_QHP} is specified as acting upon the Zeeman subspace with the projection superoperators
\begin{align}
\mathcal{P}_F &= \sum_{m_F, m'_F} \vert F, m_F; F, m'_F) ( F,m_F; F,m'_F\vert,\\
\mathcal{P}_{\mathrm{Z}} &= \mathcal{P}_a + \mathcal{P}_b.
\end{align}
The intrinsic absorption cross section in the QHP regime should be modified as 
\begin{align}
\bar{\sigma}_{\mathrm{QHP}} = A \left[ x_a \mathcal{L}(\Delta_a) + x_b \mathcal{L}(\Delta_b) \right].
\label{eq:sigma_0_QHP}
\end{align}

The corrected depopulation superoperator in Eq.~\eqref{eq:A_dp_QHP} highlights the physical consequences of the difference in absorption between the hyperfine multiplets.
Indeed, this difference is characterized by the factor $Q_F$ for $F=a$ or $F=b$ as shown in Fig.~\ref{fig2:Q_F}.
For hyperfine coherence between the hyperfine multiplets, e.g., $\vert a, m_F; b, m'_F)$, a simple form of $\mathcal{A}_{\mathrm{dp}}$ like Eq.~\eqref{eq:A_dp_QHP} is not available, as proved in Appendix~\ref{app:simplification}. 
Fortunately, the hyperfine coherence is usually neglected in analyzing most of the observables like spin polarization, magnetic-resonance linewidth, etc, as shown in Section~\ref{sec:results_and_discussion}.

\subsubsection{\label{sec:repopulation}Repopulation}

The repopulation term in Eq.~\eqref{eq:rp_def} represents the generation and relaxation of the atomic ES spin during the optical pumping process.
In a manner analogous to the treatment of $\mathcal{A}_{\mathrm{dp}}$, the optical excitation superoperator $\mathcal{A}^{\{\mathrm{eg}\}}_{\mathrm{p}}$ in the QHP regime is simplified as (see Appendix~\ref{app:simplification} for detailed proof)
\begin{align}
\mathcal{A}^{\{\mathrm{eg}\}}_{\mathrm{p}}\mathcal{P}_{\mathrm{Z}}
= 2\sum_{F} (\hat{\mathbf{e}}^{*}\cdot \mathbf{d})\otimes (\mathbf{d}^{\dagger}\cdot\mathbf{\hat{e}}) Q_{F}\mathcal{P}_F.
\label{eq:Ap_eg_QHP}
\end{align}
This result physically means that detunings $\Delta_F$ for $F=a$ and $F=b$ contribute differently to the optical excitation process.

Regarding the ES relaxation and the subsequent transition to the GS in Eq.~\eqref{eq:rp_def}, 
Appelt \textit{et al.} show that dominant mechanisms are typically $\mathrm{J}$-destruction collision and quenching \cite{Appelt1998}.
Thus $\mathcal{A}_{\mathrm{rp}}$ can be expanded in power series of $H^{\{\mathrm{e}\}\copyright}$, 
retaining only the leading term
\begin{align}
\mathcal{A}_{\mathrm{rp}} \approx \mathcal{A}^{(0)}_{\mathrm{rp}} =  \mathcal{A}_{\mathrm{q}}^{\{\mathrm{ge}\}}\left(1-\mathcal{A}_{\mathrm{JD}}\right) \mathcal{A}_{\mathrm{p}}^{\{\mathrm{eg}\}}.
\label{eq:A_rp_zeroth}
\end{align}
Under lower pressure (e.g., $0.01\text{--}0.1~\mathrm{atm}$ for $\mathrm{N}_2$),
a systematic investigation in~\cite{LancorPressure2010} offers a scale of correction from $H^{\{\mathrm{e}\}\copyright}$.
For theoretical simplicity, we still adopt the approximation of Eq.~\eqref{eq:A_rp_zeroth} in the following analysis. 
As previously researched \cite{Appelt1998,HapperBook2010},
$\mathrm{J}$-destruction collisions and quenching conserve nuclear polarization while depolarizing the ES electronic polarization.
The electric-dipole interaction similarly preserves the nuclear spin.
Because the ES hyperfine coupling has no appreciable effect on the nucleus within the short ES lifetime \cite{HapperBook2010}, as adopted in the approximation of Eq.~\eqref{eq:A_rp_zeroth}, these collisional and optical processes manipulate only the electronic part of the density matrix.
So the superoperators $\mathcal{A}_{\mathrm{dp}}$ and $\mathcal{A}_{\mathrm{rp}}$ share a common nuclear part
\begin{align}
\mathcal{A}_{\mathrm{rp}} =
\left(1-\mathcal{A}_{\mathrm{SD}}\right)\mathcal{A}_{\mathrm{dp}}.
\label{eq:rp_dp_SD}
\end{align}


\begin{figure} 
\centering 
\includegraphics[width=1\linewidth]{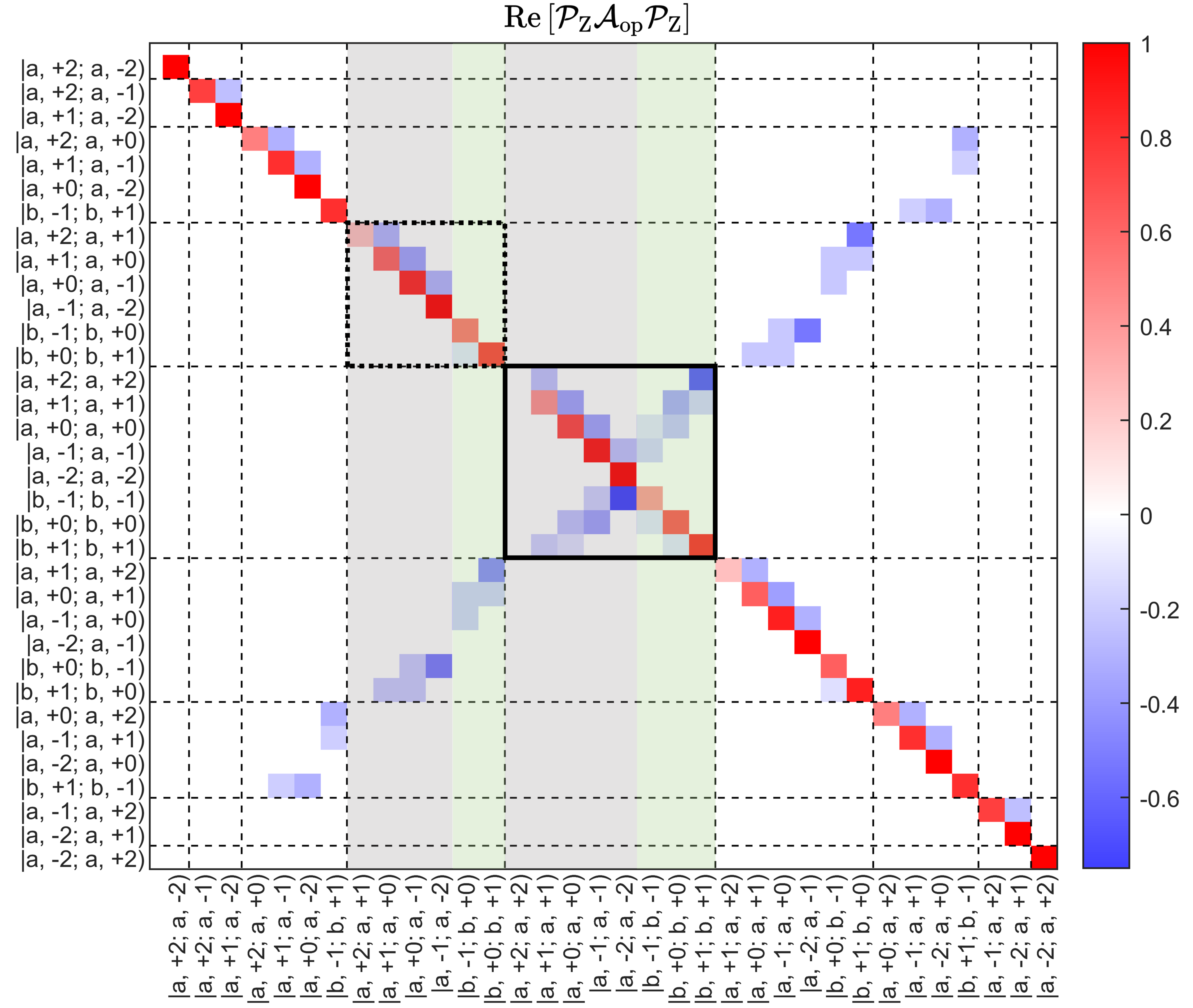} 
\caption{Real part of optical pumping superoperator within the Zeeman subspace, for nuclear spin $I=3/2$, D-line constant $K=1$, and mean photon spin $\mathbf{s}= \hat{z}$. 
The underlying matrix elements are presented in the HP limit and arranged in Liouville basis $|F,\,m_F;F,\,m'_{F})$.
They are separated by dashed lines into $m$th- to $n$th- order Zeeman coherence blocks $\mathrm{Re}\left[\mathcal{A}_{\mathrm{op}}\right]_{m,n}$ (see Appendix~\ref{app:blocks}).
The block marked by the solid lines ($m=n=0$) governs population dynamics, 
while the block marked by the dotted lines ($m=n=1$) contributes to evolution of the 1st-order Zeeman coherence.
The vertical gray and green bands encode $Q_{a}$ and $Q_b$ in Eq.~\eqref{eq:Q_F}, acting as the QHP correction. The remaining blocks follow the same structure. }
\label{fig3:A_op_QHP} 
\end{figure}

Through Eqs.~\eqref{eq:op_dp+rp}, \eqref{eq:A_dp_QHP}, and~\eqref{eq:rp_dp_SD}, 
we conclude that the full optical pumping superoperator $\mathcal{A}_{\mathrm{op}}$ with the QHP correction is expressed as
\begin{align} 
\mathcal{A}_{\mathrm{op}}\mathcal{P}_{\mathrm{Z}}&= \sum_F \left[\mathcal{A}_{\mathrm{SD}}
-\frac{K\mathbf{s}}{2}\cdot \boldsymbol{\mathcal{A}}_{\mathrm{SE}}-\frac{i\Delta_F}{\Gamma_{\mathrm{brd}}}K\mathbf{s}\cdot \mathbf{S}^{\copyright}\right]Q_F \mathcal{P}_{F}.
\label{eq:A_op_QHP} 
\end{align}
In the HP limit where $Q_F \to 1$, $\mathcal{A}_{\mathrm{op}}$ reduces to \cite{HapperBook2010}
\begin{align}
\mathcal{A}_{\mathrm{op}}^{(\mathrm{HP})} = \mathcal{A}_{\mathrm{SD}} -
\frac{K\mathbf{s}}{2}\cdot\boldsymbol{\mathcal{A}}_{\mathrm{SE}} -
\frac{i\Delta_0}{\Gamma_{\mathrm{brd}}}K\mathbf{s}\cdot\mathbf{S}^{\copyright},
\label{eq:A_op_HP}
\end{align}
where the spectral resolvability between the hyperfine multiplets vanishes.
The compact $F$-resolved form of Eq.~\eqref{eq:A_op_QHP} enables us to revisit the master equation Eq.~\eqref{eq:effective_masterEq} in the QHP regime. 
For higher computational efficiency and clearer physical interpretation in Liouville space, 
matrix elements of the superoperators are grouped into blocks according to the characteristic evolution frequencies of the density-matrix elements. 
The technical details of this procedure are provided in the Appendix~\ref{app:blocks}. 
Figure~\ref{fig3:A_op_QHP} illustrates the matrix elements and the corresponding block structure of $\mathrm{Re}\left[\mathcal{P}_{\mathrm{Z}}\mathcal{A}_{\mathrm{op}}\mathcal{P}_{\mathrm{Z}}\right]$, which represents the QHP-corrected optical pumping superoperator restricted to the Zeeman subspace of primary interest.
Specifically, the gray and green bands highlight the $Q_a$ and $Q_b$ corrections. By incorporating the asymmetric absorption strengths of the $F=a$ and $F=b$ hyperfine multiplets, these corrections to the superoperator $\mathcal{A}_{\mathrm{op}}$ provide an exact description of the resulting spin dynamics and observables.

\subsection{\label{sec:longitudinal_pumping}Observables in the quasi-high-pressure regime}
In this subsection, we investigate observables in the QHP regime 
for the common configuration in which the applied magnetic field is collinear with the pump beam. 
Specifically, we analyze the light absorption, population distributions, spin polarization, and magnetic-resonance linewidth, 
and we highlight the ways in which their behavior departs from that observed in the HP regime.

\subsubsection{\label{sec:theory_absorption}Light absorption of atoms}

Light absorption of atoms is quantified by the absorption cross section $\sigma$, 
which is expressed in terms of the effective Hamiltonian $\delta H^{\{\mathrm{g}\}}$ and the GS population $|\rho_0)$ \cite{HapperBook2010}:
\begin{align}
\sigma &= -\dfrac{2}{\Phi}\left(\mathrm{Im}\left.\left[ \delta H^{\{\mathrm{g}\}} \right]\right|\rho_0\right), 
\label{eq:sigma_GS_def}
\end{align}
where $(A|B)=\mathrm{Tr}[AB]$ denotes the inner product in Liouville space. 
Following Eqs.~\eqref{eq:delta_H_QHP}, \eqref{eq:delta_H_F}, and \eqref{eq:sigma_0_QHP}, the absorption cross section in the QHP regime is 
\begin{align}
\sigma &=  A \left[C_{a} \mathcal{L}(\Delta_a)+C_{b}  \mathcal{L}(\Delta_b)\right],
\label{eq:sigma_ab}
\end{align}
or equivalently,
\begin{align}
\sigma = \bar{\sigma}_{\mathrm{QHP}}\left(Q_{a}C_{a}+Q_{b}C_{b}\right),
\label{eq:sigma_QC}
\end{align}
with the weighting factor $C_F$ defined by
\begin{align}
C_F &= p_{F}- 2Ks\langle S_{z}\rangle_{F},
\label{eq:C_F}
\end{align}
where $s$ is the effective photon spin along the $\hat{z}$-axis ($s=0$ or $\pm 1$ correspond to linearly or circularly polarized beams), 
and $p_F$ and $\langle S_{z}\rangle_F$  are the total population and the spin polarization of the states within the hyperfine level $F$, i.e.
\begin{align}
\label{eq:p_F}
p_{F}&=\sum_{m_F} \langle F,m_{F}| \rho_{0} |F,m_{F}\rangle, \\
\label{eq:S_zF}
\langle S_{z}\rangle_F &= \sum_{m_F} \langle F,m_{F}| S_{z} |F,m_{F}\rangle\langle F,m_{F}| \rho_{0} |F,m_{F}\rangle.
\end{align}
The coefficient $C_F$ in Eq.~\eqref{eq:C_F} comprises two terms representing distinct optical pumping processes. The $F$-specific population term, $p_F$, characterizes the population redistribution between the $F=a$ and $F=b$ levels. Meanwhile, the $\langle S_z \rangle_F$ term accounts for population transfer among the $m_F$ sublevels, reflecting the generation of spin polarization through angular momentum exchange with incident photons.
Hereafter, these two mechanisms are referred to as the {\it $F$-pumping process} and {\it $m_F$-pumping process}, respectively.

In the HP limit, the two Lorentzian line shapes strongly overlap, 
such that $\mathcal{L}(\Delta_a) \approx \mathcal{L}(\Delta_b) \approx \mathcal{L}(\Delta_0)$, eliminating the $F$-resolved absorption and thereby suppressing the $F$-pumping process.
Consequently, the $C_F$ coefficient reduces to the form $C = 1 - 2Ks\langle S_z \rangle$, equivalent to the result in Refs.~\cite{HapperRMP1972, Seltzer2008},
indicating that absorption is predominantly governed by spin polarization driven by $m_F$-pumping.
In contrast, within the QHP regime, both $F$-pumping and $m_F$-pumping processes are integral to the system dynamics.
Therefore, the quantities $p_F$ and $\langle S_z \rangle_F$  
encapsulate the fundamental mechanisms governing the QHP regime.
Elucidating their dynamics as a function of pump power and frequency is crucial for uncovering the underlying principles of optical pumping.

\subsubsection{\label{sec:theory_population}Population and polarization}

In the presence of longitudinal pumping light, as well as spin-destruction and spin-exchange relaxation modeled in Eq.~\eqref{eq:effective_masterEq}, the population dynamics are governed by
\begin{align}
\frac{d\left|\rho_{0}\right)}{dt} = -\mathcal{G}_{\mathrm{pop}} \left|\rho_{0}\right),
\label{eq:pop_dynamic}
\end{align}
with the superoperator $\mathcal{G}_{\mathrm{pop}}$
\begin{align}
\mathcal{G}_{\mathrm{pop}} =& \Gamma_{\mathrm{SD}} \left[ \mathcal{A}_{\mathrm{SD}} \right]_{0,0} + R_{\mathrm{op}}\left[ \mathcal{A}_{\mathrm{op}} \right]_{0,0} \nonumber \\
&+ \Gamma_{\mathrm{EX}}\left(\left[ \mathcal{A}_{\mathrm{SD}} \right]_{0,0} -\langle S_z\rangle\left[ \mathcal{A}^{(z)}_{\mathrm{SE}} \right]_{0,0} \right),
\label{eq:G_pop}
\end{align}
where $\langle S_z\rangle=(S_{z}|\rho_{0})$ is the expectation value of the electron spin along the $z$ direction.

The steady-state solution, $\left|\rho_{\mathrm{ss}}\right) = \left|\rho_{0}(t\to+\infty)\right)$, is obtained by evaluating the null space of $\mathcal{G}_{\mathrm{pop}}$. 
Within the QHP regime, the population ratio, $\eta$, between the $|F=a, m_F\rangle$ and $|F=b, m_F\rangle$ states for a given $m_F$ ($|m_F| \neq a$) is given by
\begin{align}
\eta = \frac{(a,m_F;a,m_F\vert\rho_{\mathrm{ss}})}{(b,m_F;b,m_F\vert\rho_{\mathrm{ss}})} 
= \frac{\Gamma_{\mathrm{EX}}+\Gamma_{\mathrm{SD}}+R_{\mathrm{op}}{Q}_{b}}{\Gamma_{\mathrm{EX}}+\Gamma_{\mathrm{SD}}+R_{\mathrm{op}}{Q}_{a}}.
\end{align}
This ratio is independent of $m_F$. 
In the conventional HP limit, the absorption efficiencies become equivalent ($Q_a = Q_b$),  driving the ratio to $\eta \rightarrow 1$. 
This limit seamlessly recovers the widely recognized STD \cite{HapperSERF1977,Appelt1998}. 
However, in the scenario where $Q_a \neq Q_b$, the system diverges from this standard behavior. 
Under these conditions, the asymmetric absorption efficiencies parameterized by $Q_F$ induce an effective population transfer between the $F=a$ and $F=b$ multiplets, 
a phenomenon that dictates the core dynamics explored in the subsequent analysis.

For linearly polarized pumping with effective photon spin $s=0$, the population distribution is exactly solvable. 
Both the lack of ES resolvability and the absence of angular momentum injection preclude $m_F$-pumping.
Consequently, these magnetic sublevels remain uniformly populated, and the generation of spin polarization is strictly forbidden (i.e., $\langle S_z\rangle_F = \langle S_z\rangle\equiv 0$). 
The system dynamics therefore reduce to those of an effective two-level system comprising the hyperfine multiplets $F=a$ and $F=b$. 
Under these conditions, the analytical steady-state solution, $\left|\rho_{\mathrm{lin}}\right)$, 
dictated by the principle of detailed balance \cite{KellyMarkov2011}, is given by
\begin{align}
\left|\rho_{\mathrm{lin}}\right) &= \sum_{F,m_F}\frac{p_F}{2F+1}\left|F,m_{F};F,m_{F}\right), \\
p_F &= \frac{x_F\left(\Gamma_{\mathrm{SD}}+\Gamma_{\mathrm{EX}}+R_{\mathrm{op}}Q_{F'\neq F}\right)}{\Gamma_{\mathrm{SD}}+\Gamma_{\mathrm{EX}}+R_{\mathrm{op}}(x_aQ_{b}+x_bQ_{a})},
\label{eq:rho_lin}
\end{align}
where $\Gamma_{\mathrm{EX}}+\Gamma_{\mathrm{SD}}$ denotes the relaxation-induced population exchange rate between the multiplets, 
acting as a restorative mechanism driving the system towards equilibrium. 
Conversely, $R_{\mathrm{op}}Q_{F'\neq F}$ characterizes the effective population transfer from the $F'$ to the $F$ multiplet,
creating unbalanced population between them under the linearly polarized pumping.

\begin{figure}
\centering 
\includegraphics[width=1\linewidth]{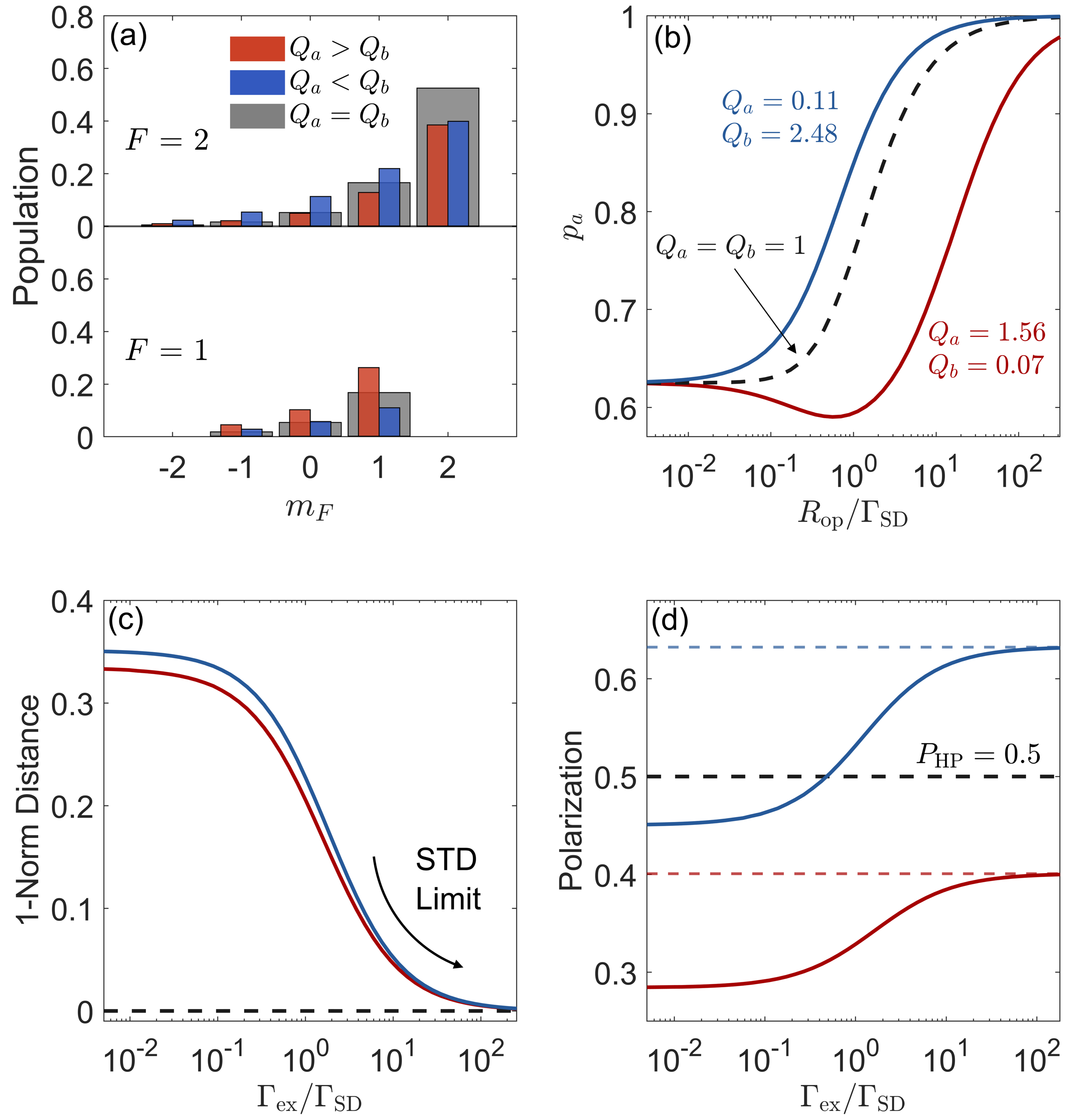} 
\caption{(a) Representative steady-state populations under slow spin-exchange conditions. 
(b) Total population of the $F=a$ multiplet as a function of the normalized pumping rate $R_{\mathrm{op}}/\Gamma_{\mathrm{SD}}$, with curves denoting different $Q_F$ configurations. 
(c) The 1-norm distance $d$ [see Eq.~\eqref{eq:d}] between the steady-state population and an STD profile of the equivalent polarization, fixing $R_{\mathrm{op}}/\Gamma_{\mathrm{SD}}=1$. 
(d) Spin polarization $P$ versus $\Gamma_{\mathrm{EX}}/\Gamma_{\mathrm{SD}}$. 
Color coding for (c) and (d) is consistent with (b). Results are obtained for $I=3/2$ and $K=s=1$.}
\label{fig4:spin_temperature_and_population} 
\end{figure}
Under circularly polarized pumping ($s=\pm 1$), the system dynamics become increasingly complex due to the concurrent action of $F$-pumping and $m_F$-pumping.
Because these coupled processes are governed by the nonlinear Eq.~\eqref{eq:pop_dynamic}, which lacks an analytical solution, the steady-state distributions are solved via numerical simulation.

Specifically, in the standard HP limit, solely $m_F$-pumping dictates the dynamics;
this mechanism drives the population toward states with higher $m_F$ values while maintaining the population balance between the $F=a$ and $F=b$ multiplets for a common $m_F$.
Conversely, within the QHP regime, the emergence of $F$-pumping breaks this balance.
As illustrated in Fig.~\ref{fig4:spin_temperature_and_population}(a), when $Q_a > Q_b$, the disproportionate absorption within the $F=a$ multiplet pumps the population into the $F=b$ levels (red bars).
An opposite redistribution trend occurs under the $Q_b > Q_a$ condition (blue bars). 
Figure~\ref{fig4:spin_temperature_and_population}(b) elucidates the dependence of this impact on the normalized pumping intensity, parameterized as $R_{\mathrm{op}}/\Gamma_{\mathrm{SD}}$. 
This redistribution mechanism exerts a negligible influence 
when approaching either the weak-pumping limit ($R_{\mathrm{op}} \ll \Gamma_{\mathrm{SD}}$) of a nearly uniform distribution or the strong-pumping limit ($R_{\mathrm{op}} \gg \Gamma_{\mathrm{SD}}$) of stretched-state saturation. 
In contrast, the intermediate domain clearly manifests the $F$-pumping dynamics driven by $Q_a \neq Q_b$, where, for instance, the notably suppressed red curve (compared with the dashed line) illustrates the net transfer from the $F=a$ to the $F=b$ levels.

The $F$-pumping-induced imbalance in the QHP regime drives the system away from the conventional STD characteristic of the HP limit.
Specifically, the steady-state density matrix can no longer be parameterized by a single temperature factor $\beta(P)$ in the form $|\rho_{\mathrm{STD}}) \propto \exp[\beta(P) F_z]$ \cite{HapperSERF1977,Appelt1998}, 
with $F_z=S_z+I_z$ being the $z$-component of the total angular momentum operator ($I_z$ the nuclear spin counterpart), and $P=2(S_{z}|\rho_{\mathrm{ss}})$ being the steady-state spin polarization. 
To quantify this deviation, we define a 1-norm distance metric
\begin{align}
d = \Vert \rho_{\mathrm{ss}} -  \rho_{\mathrm{STD}}\Vert_1 
\label{eq:d}
\end{align} 
which compares the QHP steady-state distribution to an ideal STD possessing an identical spin polarization $P$.
Figure~\ref{fig4:spin_temperature_and_population}(c) plots this distance as a function of $\Gamma_{\mathrm{EX}} / \Gamma_{\mathrm{SD}}$—the ratio of the spin-exchange rate to the spin-destruction rate—at a fixed optical pumping rate $R_{\mathrm{op}}$.
The distance $d$ decreases monotonically as the ratio increases. 
This demonstrates that rapid spin-exchange collisions provide a dominant restorative mechanism, 
effectively thermalizing the system to counteract the $F$-pumping induced by asymmetric absorption. 
The high validity of the STD approximation under the strong spin-exchange condition ($\Gamma_{\mathrm{EX}} / \Gamma_{\mathrm{SD}}\gg 1$) establishes a critical theoretical foundation 
for utilizing the STD to simplify the physical models of various practical applications, such as SERF magnetometers, even when the HP condition is not necessarily fully satisfied.

Although the STD serves as a valid approximation at high spin-exchange rate, the joint contribution of $F$- and $m_F$-pumping under QHP condition gives rise to a spin polarization different from that obtained in the HP condition.
In the HP limit with $Q_a=Q_b$ [see the black dashed line in Fig.~\ref{fig4:spin_temperature_and_population}(d)], the spin polarization reduces to the widely used form \cite{HapperRMP1972, Seltzer2008, HapperSERF1977}
\begin{align}
P_{\mathrm{HP}} = \frac{KsR_{\mathrm{op}} }{ R_{\mathrm{op}} + \Gamma_{\mathrm{SD}}}.
\label{eq:P_HP}
\end{align}
In contrast, under QHP conditions (blue and red dashed lines), the asymptotic polarization at a high spin-exchange rate remains to be sensitive to the asymmetric absorption strengths ($Q_a$ and $Q_b$).
Consequently, within the QHP framework, the spin polarization $P$ under the STD approximation cannot be obtained trivially 
and must be reshaped by the imbalance $Q_a\neq Q_b$.
Detailed results and their corresponding observable effects will be presented in Section~\ref{sec:results_and_discussion}.

\subsubsection{\label{sec:theory_magnetic_resonance}Intrinsic magnetic resonance}

The population imbalance across the $m_F$ sublevels enables Zeeman magnetic resonance under 
a radio-frequency (RF) driving field oscillating at $\omega_{\mathrm{rf}}$. 
To streamline the theoretical analysis, we conceptualize the framework established in our previous work \cite{Tang2025} under the notion of \textit{intrinsic} magnetic resonance.
This intrinsic regime describes the 1st-order Zeeman coherence generated by an infinitesimally weak driving field, 
operating under a weak static magnetic field where the nonlinear Zeeman effect is negligible, so that the spin precession frequencies degenerate into a single Larmor frequency $\omega_{\mathrm{L}}$.
In this case, the resonance dynamics are captured by the generated 1st-order Zeeman coherence $\left|\tilde{\rho}_{1}\right)$.  
In the RF rotating frame, the coherence is governed by
\begin{align}
\left|\tilde{\rho}_{1}\right) \sim \Omega_{\mathrm{R}}\mathcal{G}^{-1}_{\mathrm{WDA}}\big[S^{\copyright}_{x}\big]_{1,0}\left|{\rho}_{\mathrm{ss}}\right), 
\end{align}
where $\Omega_{\mathrm{R}}$ is the weak Rabi frequency, $\big[S^{\copyright}_{x}\big]_{1,0}$ dictates the coherence generation from polarized atoms, and $\mathcal{G}_{\mathrm{WDA}}$ is the total relaxation superoperator under the weak driving approximation. 
Since the resonance is measured by observing the transverse spin component via Faraday rotation \cite{HapperProbe1967,BudkerOM2007}, the signal is characterized by the complex amplitude
\begin{align}
R=(S_{x}\vert\tilde{\rho}_{1}) &\approx \frac{\Omega_{\mathrm{R}}}{2} \frac{W}{\Delta_{\mathrm{rf}}+i\Gamma_{2}}.
\label{eq:S_def}
\end{align}
Here, $W$ represents the dominant Zeeman coherence weight, and $\Delta_{\mathrm{rf}} = \omega_{\mathrm{rf}}-\omega_{\mathrm{L}}-\delta\omega_{\mathrm{L}}$ defines the effective RF detuning, taking into account the optical shift $\delta\omega_{\mathrm{L}}$. 
The linewidth $\Gamma_2$ is obtained by the minimal eigenvalue of $\mathcal{G}_{\mathrm{WDA}}$, and the optical shift originates from the value of $\mathrm{Im}[\mathcal{A}_{\mathrm{op}}]_{1,1}$ \cite{Tang2025}.
Experimentally, properties of the resonance directly manifest as the in-phase and out-of-phase quadratures, i.e.
\begin{align}
X&=\mathrm{Re}[R], \quad Y=-\mathrm{Im}[R].
\label{eq:XY}
\end{align}

Transitioning this framework into the QHP regime necessitates a $Q_F$-induced modification.
While all other relaxation mechanisms in $\mathcal{G}_{\mathrm{WDA}}$ remain formally unchanged, 
the optical pumping contribution to $\left|\tilde{\rho}_{1}\right)$ relaxation is replaced by the hyperfine-resolved superoperator $\left[\mathcal{A}_{\mathrm{op}}\right]_{1,1}$ 
[see Eq.~\eqref{eq:A_op_QHP} \& Fig.~\ref{fig3:A_op_QHP}] to incorporate the $Q_a \neq Q_b$ absorption asymmetry.
Detailed results about the linewidth calculations and the magnetic resonance enhancement under QHP conditions are discussed in Sec.~\ref{sec:magnetic_resonance_enhancement}.

\section{\label{sec:results_and_discussion}RESULTS AND DISCUSSION}

While the theoretical framework developed in the preceding sections is generally applicable to both the D$_1$ and D$_2$ transitions of alkali-metal atoms, the following analysis of observables in the QHP regime focuses specifically on the D$_1$ line ($K=1$). This choice is motivated by its widespread adoption in practical atomic magnetometers, which fundamentally stems from its higher optical pumping efficiency resulting from its distinct coupled energy-level structure \cite{HapperRMP1972,Seltzer2008}. Detailed comparative results for the D$_2$ line are given in Appendix~\ref{app:D2_pumping}.

\subsection{\label{sec:light_absorption_cross_section}Absorption cross section}

For optically pumped atoms in the QHP regime, 
we utilize the description of light absorption in Sec.~\ref{sec:theory_absorption} 
to calculate the absorption cross section $\sigma_{\mathrm{lin}}$ and $\sigma_{\mathrm{cir}}$ for linearly- and circularly-polarized laser beams. 

\begin{figure}[]
\centering
\includegraphics[width=1\linewidth]{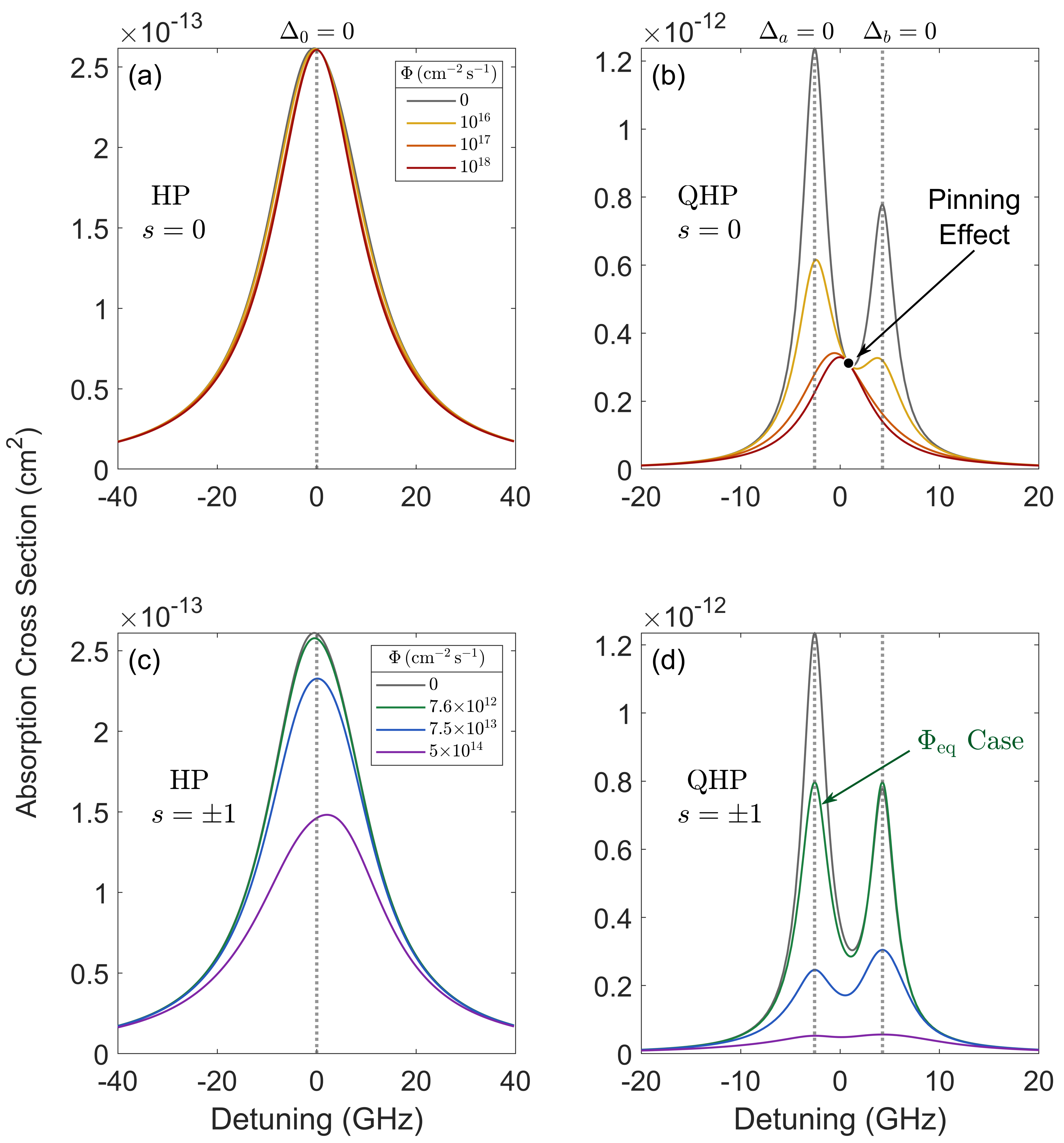}
\caption{
Absorption cross section versus light detuning for the D$_1$ line pumping. (a)--(b) Results for linearly polarized pumping ($s=0$), under typical HP ($2\Gamma_{\mathrm{brd}}=20~\mathrm{GHz}$) and QHP ($2\Gamma_{\mathrm{brd}}=3~\mathrm{GHz}$) conditions. The black marker in (b) illustrates the pinning effect at the central detuning $\Delta_{\mathrm{c}}$ in Eq.~\eqref{eq:delta_c}.
(c)--(d) Analogous plots for circularly polarized pumping ($s=\pm 1$). The $\Phi_{\mathrm{eq}}$ case, i.e., the green solid curve for which the two absorption peaks are of equal height, is labeled in (d).
In each panel, the vertical dotted lines denote the main resonance centers.
The $^{87}$Rb atoms are considered for simulations, with a fixed $\Gamma_{\mathrm{EX}}=7.75~\mathrm{kHz}$. To account for the spin-destruction collision induced by buffer gas, $\Gamma_{\mathrm{SD}}$ is set to $0.05~\mathrm{kHz}$ and $0.27~\mathrm{kHz}$ for the QHP and HP cases.
}
\label{fig5:absorption_cross_section} 
\end{figure}

The absorption cross section for a linearly polarized laser beam is given by
\begin{align}
\sigma_{\mathrm{lin}} &=  A\left[p_a \mathcal{L}(\Delta_a) + p_b \mathcal{L}(\Delta_b)\right],
\end{align}
where the populations $p_a$ and $p_b$ for the respective multiplets are derived from Eq.~\eqref{eq:rho_lin}.
Figures~\ref{fig5:absorption_cross_section}(a) and \ref{fig5:absorption_cross_section}(b) illustrate the absorption cross sections within the HP and QHP regimes. 
In the standard HP limit [Fig.~\ref{fig5:absorption_cross_section}(a)], the absorption profile is remarkably insensitive to the incident laser power, 
exhibiting a negligible optical pumping effect. 
This behavior is physically consistent: because linearly polarized photons transfer no net angular momentum to the electron spins, 
the GS population distribution remains largely unperturbed.
However, within the QHP regime [Fig.~\ref{fig5:absorption_cross_section}(b)], 
the reduced collisional broadening enables the two distinct hyperfine transitions to be clearly resolved. 
Crucially, the absorption cross section experiences severe suppression when the excitation frequency is tuned near resonance with either the $F=a$ or $F=b$ levels.
This suppression becomes increasingly pronounced as the incident photon flux, $\Phi$, increases. 
In the strong-flux limit, the two well-resolved peaks collapse and merge into a significantly weaker, broadened profile.

The absorption suppression shown in Fig.~\ref{fig5:absorption_cross_section}(b) arises from the $F$-pumping process in the QHP regime. 
The pumping near resonance ($\Delta_F = 0$) efficiently transfers the population from a driven hyperfine multiplet $F$ into the alternate multiplet $F' \neq F$, 
thereby suppressing the overall absorption. 
As the incident photon flux $\Phi$ increases, this $F$-pumping mechanism intensifies, 
eventually driving the atomic population almost entirely into a single hyperfine multiplet. 
Consequently, in the strong-pumping limit ($R_{\mathrm{op}} \gg \Gamma_{\mathrm{SD}} + \Gamma_{\mathrm{EX}}$), 
as long as the condition of negligible population trapping in the ES still holds, 
the absorption cross section asymptotically converges to the analytical form
\begin{align}
\sigma_{\mathrm{lin}}(R_{\mathrm{op}}\gg \Gamma_{\mathrm{SD}}+\Gamma_{\mathrm{EX}}) &= \frac{A_{\mathrm{eff}}}{\pi} \frac{ \Gamma_{\mathrm{eff}} }{\Delta^{2}_{0}+\Gamma^{2}_{\mathrm{eff}}}.
\label{eq:sigma_lin_limit}
\end{align}
This expression characterizes a single Lorentzian profile centered at the reference frequency $\Delta_0 = x_a\omega^{\{\mathrm{g}\}}_{a,0} + x_b\omega^{\{\mathrm{g}\}}_{b,0} = 0$.
Here, the effective spectral area and linewidth are modified to $A_{\mathrm{eff}} = A/\xi$ and $\Gamma_{\mathrm{eff}} = \xi\Gamma_{\mathrm{brd}}$, respectively,
where the dimensionless scaling factor $\xi$ is given by
\begin{align}
\xi &= \sqrt{1+x_a x_b\left(\frac{\omega^{\{\mathrm{g}\}}_{\mathrm{hf}}}{\Gamma_{\mathrm{brd}}}\right)^{2}},
\label{eq:xi}
\end{align}
with $\omega^{\{\mathrm{g}\}}_{\mathrm{hf}}$ denoting the GS hyperfine splitting.

As the incident photon flux $\Phi$ is varied, an interesting {\it pinning effect} emerges: 
a unique intensity-invariant crossing point at the central detuning [Fig.~\ref{fig5:absorption_cross_section}(b)]
\begin{align}
\Delta_{\mathrm{c}} = \frac{1}{2}\left(\omega^{\{\mathrm{g}\}}_{a,0} + \omega^{\{\mathrm{g}\}}_{b,0}\right),
\label{eq:delta_c}
\end{align}
where the absorption cross section $\sigma_{\mathrm{lin}}$ remains strictly independent of $\Phi$.
This characteristic point arises because the pump light is equally detuned from both hyperfine multiplets, yielding $Q_a = Q_b$.
This symmetric absorption strength effectively suppresses the $F$-pumping dynamics entirely, 
pinning the cross section at the constant value $\bar{\sigma}_{\mathrm{QHP}}$ defined in Eq.~\eqref{eq:sigma_0_QHP}.

For the absorption cross section under circularly polarized pumping ($s=\pm 1$), by invoking the STD approximation detailed in Sec.~\ref{sec:theory_population}, 
we substitute the analytical expressions for $p_{F}$ and $\langle S_{z}\rangle_{F}$ (summarized in Table~\ref{tab:p_F_S_zF}) into Eq.~\eqref{eq:C_F}. 
This yields the total absorption cross section:
\begin{align}
\sigma_{\mathrm{cir}} &= A\left[C_a(I,\,P) \mathcal{L}(\Delta_a) + C_b(I,\,P) \mathcal{L}(\Delta_b)\right],
\end{align}
where the coefficients $C_F$ are now explicitly parameterized by the nuclear spin $I$ and the spin polarization $P$. Taking an isotope with $I=3/2$ as an example, these coefficients take the exact forms:
\begin{align}
C_{a=2} &= \frac{5}{8} - \frac{10P-5P^2+6P^3-P^4}{8(1+P^2)}, \\
C_{b=1} &= \frac{3}{8} + \frac{2P-5P^2-2P^3-P^4}{8(1+P^2)}.
\end{align}

The corresponding theoretical spectra under varying incident photon fluxes, $\Phi$, are plotted for the HP and QHP regimes in Figs.~\ref{fig5:absorption_cross_section}(c) and \ref{fig5:absorption_cross_section}(d), respectively. Unlike the linearly polarized scenario, circularly polarized light effectively transfers net angular momentum to the atomic system. As $\Phi$ increases, the atoms are continuously pumped into the "dark" stretched state ($|F=a, m_F=sa\rangle$), where they can no longer absorb incident photons. This physical mechanism manifests as a pronounced, global reduction in the absorption cross section across both regimes. In the HP limit [Fig.~\ref{fig5:absorption_cross_section}(c)], strong collisional broadening smears the hyperfine structure into a single unresolvable peak, which undergoes overall amplitude suppression as $P$ increases. Conversely, in the QHP regime [Fig.~\ref{fig5:absorption_cross_section}(d)], the reduced broadening allows the two hyperfine transitions to remain distinctly resolved. Nevertheless, driven by the same global accumulation of population in the dark stretched state, both resolved peaks experience simultaneous and significant absorption suppression.

\begin{figure}[t]
\includegraphics[width=1\linewidth]{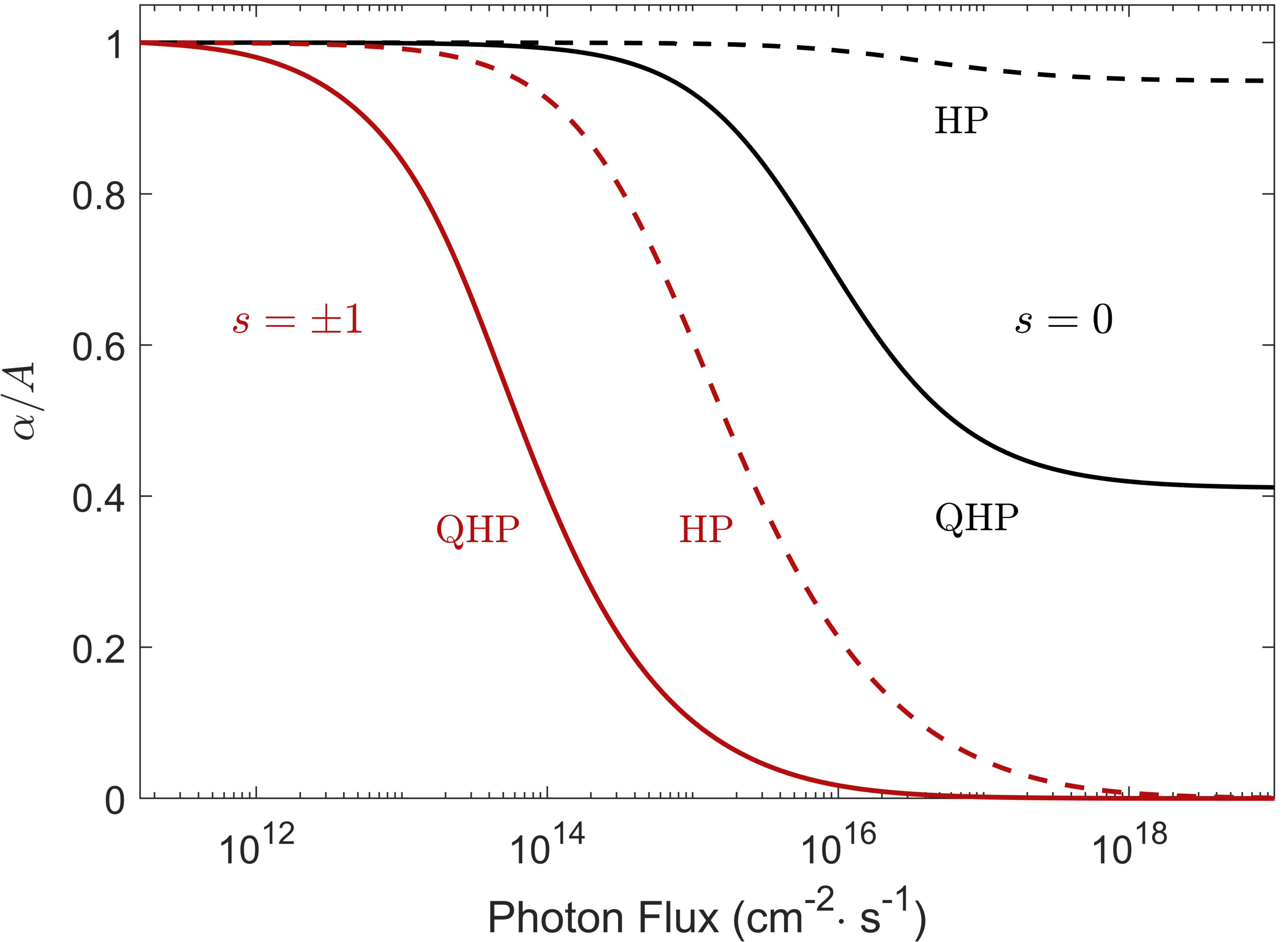} 
\caption{
Integrated spectral area $\alpha$ [see Eq.~\eqref{eq:alpha}] normalized by $A$ versus photon flux $\Phi$, incorporating both linearly ($s=0$) and circularly ($s=\pm 1$, D$_1$ line) polarized pumping. Solid and dashed lines correspond to the QHP and HP cases, respectively. With increasing $\Phi$, the $s=0$ curves asymptotically converge to $1/\xi=0.41$ and $0.95$ [see Eq.~\eqref{eq:xi}]. Parameters are consistent with Fig.~\ref{fig5:absorption_cross_section}.
}
\label{fig6:integrated_sigma} 
\end{figure}

As a secondary observation within the QHP regime, Fig.~\ref{fig5:absorption_cross_section}(d) reveals an intensity-dependent peak inversion: 
the dominant resonance shifts from $\Delta_a = 0$ at low photon flux to $\Delta_b = 0$ at high flux.
The threshold flux equalizing these peaks corresponds to the green solid curve, and coincides with the photon flux $\Phi_{\mathrm{eq}}$ discussed in Sec.~\ref{sec:spin_polarization}.
This crossover, occurring when $C_a = C_b$, is driven by competing mechanisms.
At low $\Phi$, the unpolarized absorption is primarily dictated by the higher state degeneracy of the $F=a$ multiplet.
However, as macroscopic polarization builds, the population shifts into specific magnetic sublevels,
causing the $F=b$ multiplet—which inherently possesses a larger fraction of active electronic spin components—to dominate the absorption process.
The specific polarization dynamics at this equalization point are addressed in the subsequent section.

To evaluate the macroscopic absorption capacity independent of specific laser detuning, 
we define the integrated spectral area:
\begin{align}
\alpha = \int \sigma(\Delta_0) \, d\Delta_0.
\label{eq:alpha}
\end{align}
Figure~\ref{fig6:integrated_sigma} illustrates the dependence of the normalized integrated area on the incident photon flux 
for both linear and circular polarizations across the HP and QHP regimes.
Under linearly polarized pumping (black curves), $\alpha$ monotonically decays from its initial value $A$ 
and asymptotically plateaus at an effective area $A_{\mathrm{eff}}$ in the strong-pumping limit, 
perfectly corroborating Eq.~\eqref{eq:sigma_lin_limit}. 
Crucially, the QHP regime exhibits a pronounced reduction in absorption (solid black line), 
whereas the HP case demonstrates almost complete resilience to optical bleaching (dashed black line). 
This stark contrast is governed by the scaling factor $\xi$ defined in Eq.~\eqref{eq:xi}. 
Conversely, circularly polarized pumping (red curves) systematically drives $\alpha$ to zero.
This total suppression occurs because continuous angular momentum transfer via $m_F$-pumping perfectly polarizes the atomic ensemble into the non-absorbing stretched state, 
ultimately rendering the medium fully transparent. 
This complete optical bleaching is achieved at significantly lower photon fluxes under QHP conditions compared to the HP case, 
a direct consequence of the stronger resonant absorption enabled by the resolved hyperfine structure.

\begin{table}[]
\caption{
Analytic expressions for $p_a$ and $\langle S_z \rangle_a$ defined in Eqs.~\eqref{eq:p_F} and \eqref{eq:S_zF}. These relations are derived under the STD approximation, as a function of nuclear spin $I$ and the spin polarization $P$. 
The $F=b$ counterparts follow as $p_{b}=1-p_{a}$ and $\langle S_{z}\rangle_{b}= P/2 -\langle S_{z}\rangle_{a}$. 
The rightmost column lists the equal-polarization point $P_{\mathrm{eq}}$ for D$_1$ line pumping, obtained by solving Eq.~\eqref{eq:P_eq} with $K=|s|=1$. For $I > 1/2$, these results exclude the trivial $\vert P\vert=1$ root.
}
\centering
\renewcommand{\arraystretch}{1.25}
\begin{ruledtabular}
\begin{tabular}{cccc}
$I$ & $p_{a}(I,P)$ & $\langle S_{z}\rangle_{a}(I,P)$ & $\vert P_{\mathrm{eq}}\vert$ \\
\hline \\[-10pt]
$\dfrac{1}{2}$ &
$\dfrac{3}{4} + \dfrac{ P^{2}}{4}$ &
$\dfrac{P}{2}$ & 1 \\[6pt]

$1$ &
$\dfrac{2}{3} + \dfrac{4P^2}{3(3+ P^2)}$ &
$\dfrac{5P+P^{3}}{3(3+P^{2})}$ & 0.394 \\[6pt]

$\dfrac{3}{2}$ &
$\dfrac{5}{8} + \dfrac{5P^2 + P^4}{8(1 + P^2)}$ &
$\dfrac{5P + 3 P^{3}}{8\left(1 + P^{2}\right)}$ & 0.207 \\[6pt]

$2$ &
$\dfrac{3}{5} + \dfrac{4(5P^2 + 3P^4) }{5(5 + 10 P^2 + P^4)}$ &
$\dfrac{35P+42P^3 +3P^5 }{10(5 + 10 P^2 + P^4)}$ & 0.128 \\[6pt]

$\dfrac{5}{2}$ &
$\dfrac{7}{12} + \dfrac{35P^2 + 42 P^4 + 3 P^6}{12\left(3 + 10 P^{2} + 3 P^{4}\right)}$ &
$\dfrac{7P + 14 P^{3} + 3 P^{5}}{3\left(3 + 10 P^{2} + 3 P^{4}\right)}$ & 0.087
\end{tabular}
\end{ruledtabular}

\label{tab:p_F_S_zF}
\end{table}
\subsection{\label{sec:spin_polarization}Spin polarization}
\begin{figure}
\includegraphics[width=0.9\linewidth]{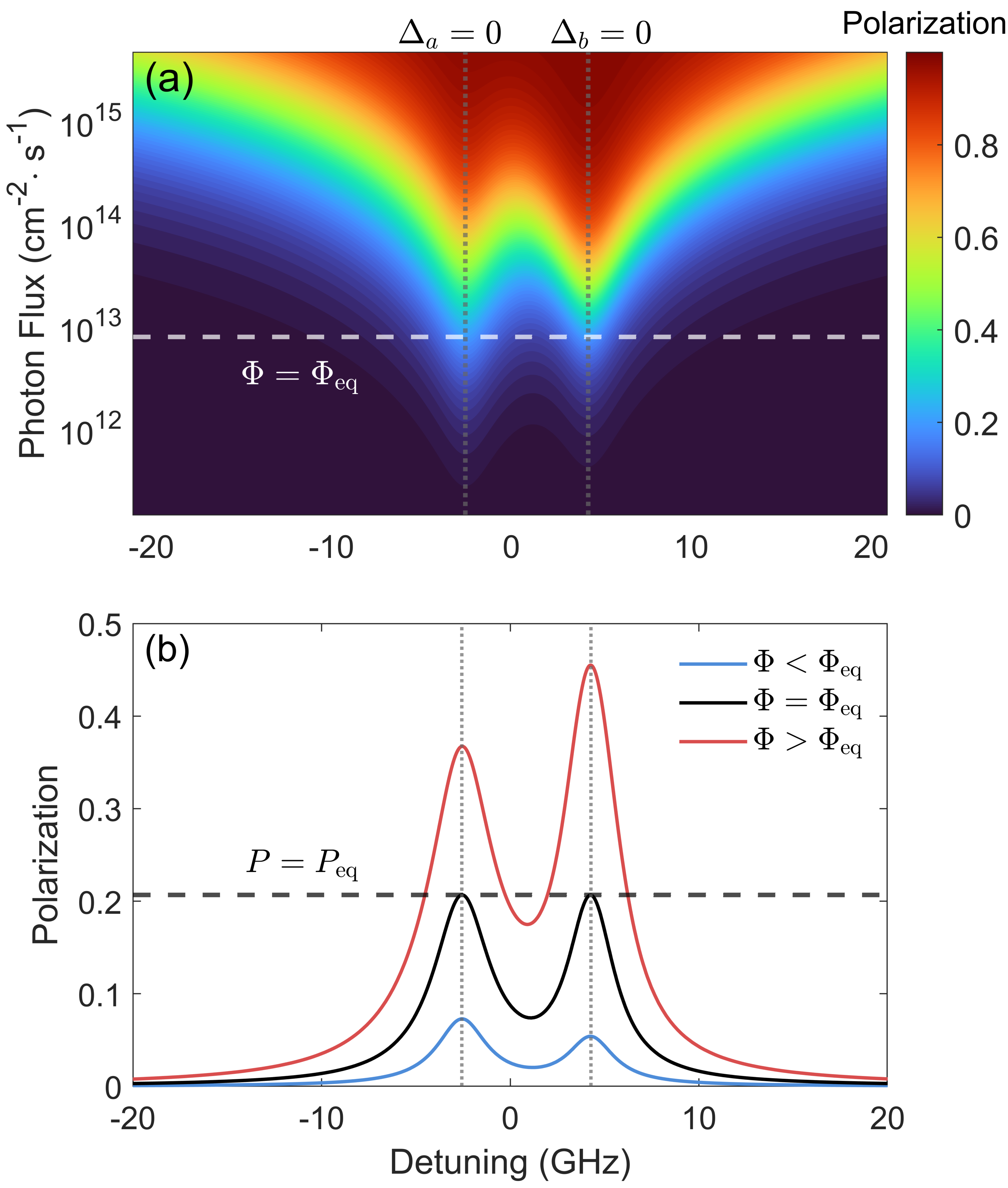} 
\caption{
(a) Two-dimensional color map of spin polarization $P$, with $K=s=1$ and $2\Gamma_{\mathrm{brd}}=3$ GHz. 
The white line signifies the photon flux $\Phi_{\mathrm{eq}}=7.6 \times 10^{12}~\mathrm{cm^{-2} s^{-1}}$ that leads to the equal-polarization point. 
(b) Variation of polarization with light detuning at representative $\Phi$ values. The black dashed line isolates the $\Phi_{\mathrm{eq}}$ case, alongside the solid blue and red curves for $\Phi = 2 \times10^{12}$ and $2 \times 10^{13}~\mathrm{cm^{-2} s^{-1}}$. The spin-destruction rate for $^{87}$Rb atoms used here is fixed at $\Gamma_{\mathrm{SD}}=4.7~\mathrm{Hz}$.
}
\label{fig7:polarization_sweeper} 
\end{figure}

Building upon the population dynamics established in Sec.~\ref{sec:theory_population}, 
we now evaluate the steady-state spin polarization $P$ within the QHP regime. 
Projecting the observable $(F_{z}|=(I_{z}|+(S_{z}|$ onto both sides of Eq.~\eqref{eq:pop_dynamic} yields the evolution equation for the average total angular momentum $\langle F_{z} \rangle$:
\begin{align}
\frac{d\langle F_{z}\rangle}{dt} &=  \frac{1}{2}R_{\mathrm{op}} \sum_{F}{Q}_{F}\left( sp_{F} -2\langle S_{z}\rangle_F \right) -\Gamma_{\mathrm{SD}}\langle S_{z}\rangle.
\label{eq:F_z_equation}
\end{align}
Enforcing the steady-state condition $d\langle F_{z}\rangle/dt=0$ provides an implicit equation for the macroscopic polarization:
\begin{align}
P &= \frac{sR_{\mathrm{op}}({Q}_{a}p_{a}+{Q}_{b}p_{b})} {R_{\mathrm{op}}\left({Q}_{a}u_{a}+{Q}_{b}u_{b}\right)+\Gamma_{\mathrm{SD}}},
\label{eq:spin_polarization}
\end{align}
where $u_F=\left\langle S_{z}\right\rangle_F/\left\langle S_{z}\right\rangle$ denotes the relative polarization weight of the $F$ multiplet.
Combining the steady-state solution of Eq.~\eqref{eq:F_z_equation} with
$\sigma_{\mathrm{cir}}$ Eq.~\eqref{eq:sigma_ab} yields a direct link
between light absorption and spin polarization for $s=\pm 1$,
\begin{align}
P = \frac{s\Phi\sigma_{\mathrm{cir}}}{\Gamma_{\mathrm{SD}}},
\label{eq:P_sigma}
\end{align}
which is useful for the measurement and calibration of the polarization
in practical experiments.
While Eqs.~\eqref{eq:F_z_equation}, \eqref{eq:spin_polarization} and \eqref{eq:P_sigma} are mathematically exact and structurally parallel to those derived for the HP regime in Ref.~\cite{Appelt1998},
a fundamental distinction defines the QHP regime.
Specifically, the dynamics are inherently modulated by the parameters $Q_a$ and $Q_b$,
which strictly account for the asymmetric absorption efficiencies between the $F=a$ and $F=b$ multiplets.

By employing the STD approximation,
we substitute the explicit expressions from Table~\ref{tab:p_F_S_zF} to convert Eq.~\eqref{eq:spin_polarization} into a polynomial equation in $P$,
which is readily solved for a given set of rate parameters.
Under typical QHP conditions, the resulting steady-state polarization $P$ exhibits several distinct features, as illustrated in Fig.~\ref{fig7:polarization_sweeper}.
Figure~\ref{fig7:polarization_sweeper}(a) demonstrates that $P$ generally accumulates with increasing $\Phi$,
manifesting two distinct maxima at the resonances $\Delta_a=0$ and $\Delta_b=0$ where the hyperfine structure is well-resolved.
Figure~\ref{fig7:polarization_sweeper}(b), which profiles representative slices across different $\Phi$ levels, highlights an intensity-dependent shift in pumping dominance.
Specifically, lower photon fluxes favor polarization via the $\Delta_a=0$ resonance (blue curve), whereas higher fluxes render the $\Delta_b=0$ resonance more efficient (red curve).
This behavior directly mirrors the dynamics of the D$_1$ absorption cross section $\sigma_{\mathrm{cir}}$ discussed in Sec.~\ref{sec:light_absorption_cross_section}:
stronger resonant absorption induces a higher rate of angular momentum transfer, consequently yielding a higher macroscopic $P$.

\begin{figure*}[t]
\centering 
\includegraphics[width=0.78\linewidth]{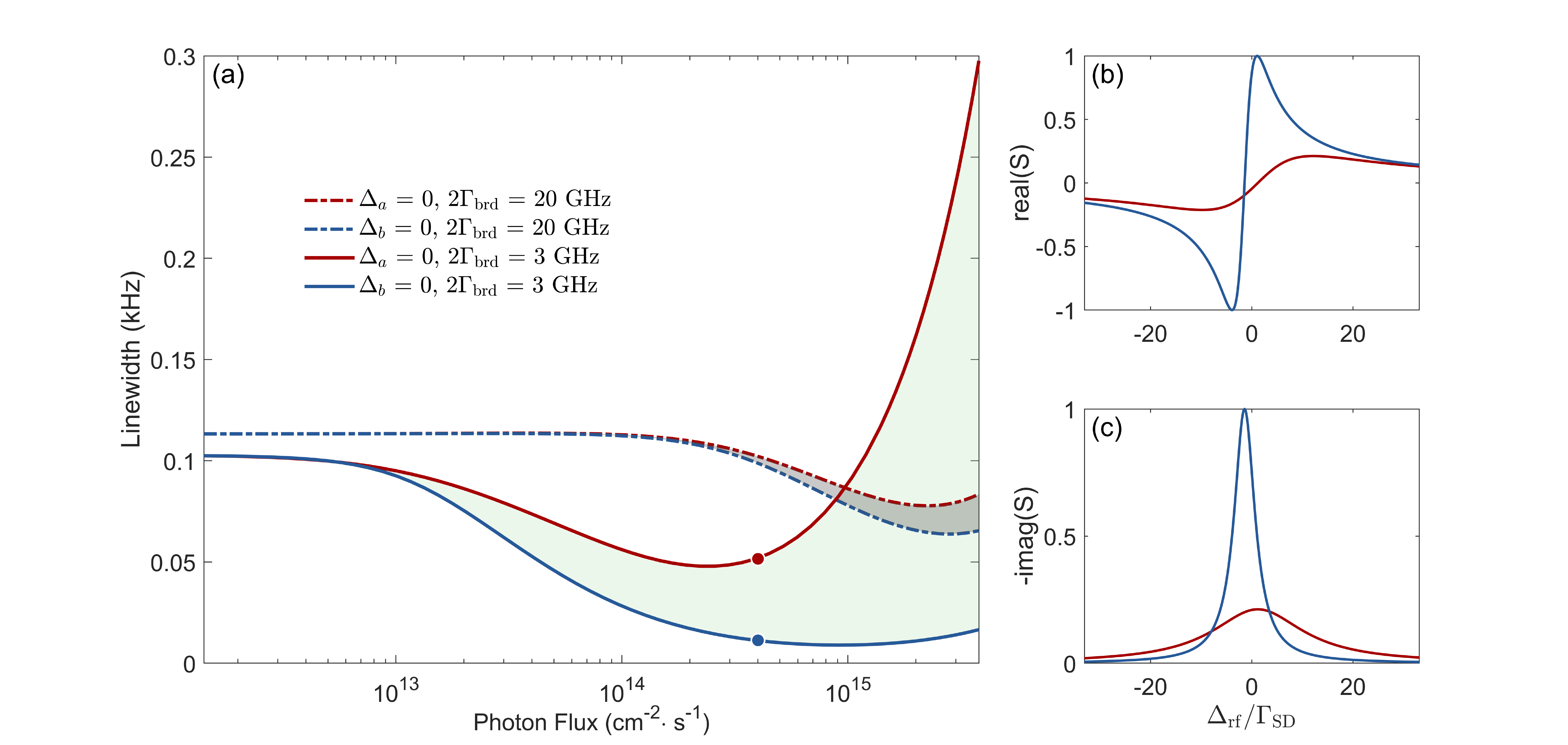}
\caption{
(a) Magnetic-resonance linewidth $\Gamma_2$ evaluated at D$_1$ line resonant pumping. The solid and dashed curves denote the QHP and HP cases. The shaded regions highlight the narrowing effect when the pumping light targets the $F=b$ level in the QHP regime. 
Panels (b) and (c) display the normalized in-phase and out-of-phase quadratures ($X$ and $Y$) of the complex response $R$ [see Eq.~\eqref{eq:XY}]. The red, blue, and black dashed curves represent $\Delta_a=0$, $\Delta_b=0$ (QHP), and the HP case, respectively, at the photon flux $\Phi = 4\times 10^{14}~\mathrm{cm^{-2} s^{-1}}$ marked in (a). For these calculations, we consider $^{87}$Rb atoms subjected to a weak magnetic field $B=5~\mu\mathrm{T}$. 
Adopting consistent relaxation parameters in this section, the linewidth determined entirely by collision rates [see Eq.~\eqref{eq:Gamma2_baseline}] is set to $0.10$~kHz for the QHP case and $0.11$~kHz for the HP case.
This slight difference accounts for the variation in $\Gamma_{\mathrm{SD}}$ induced by the respective buffer gas pressures.
}
\label{fig8:linewidth_narrowing} 
\end{figure*}

As a supplementary observation of interest, a critical equal-polarization point emerges at the equal-absorption ($C_a=C_b$) photon flux $\Phi_{\mathrm{eq}}$ for $s=\pm 1$, 
where the polarization amplitudes at the $F=a$ and $F=b$ resonances equalize to a common value, $P_{\mathrm{eq}}$.
This condition is demarcated by the white dashed line in Fig.~\ref{fig7:polarization_sweeper}(a) and the solid black curve in Fig.~\ref{fig7:polarization_sweeper}(b).
Based on Eq.~\eqref{eq:spin_polarization}, extracting this critical condition is mathematically equivalent to enforcing
\begin{align}
sp_a - 2\langle S_z \rangle_a &= sp_b - 2\langle S_z \rangle_b,
\label{eq:P_eq}
\end{align}
which corresponds to a required photon flux of
\begin{align}
\Phi_{\mathrm{eq}} &= \frac{ P_{\mathrm{eq}}\Gamma_{\mathrm{SD}}}{A(sp_F - 2\langle S_z \rangle_F) [\mathcal{L}(0) + \mathcal{L}(\omega^{\{\mathrm{g}\}}_{\mathrm{hf}})] }.
\label{eq:Phi_eq}
\end{align}
Equation~\eqref{eq:P_eq} represents a fundamental balance independent of collisional broadening or relaxation rates.
Physically, it indicates that the combined mechanisms of $F$- and $m_F$-pumping exert an identical net impact on $P$ via either the $F=a$ or $F=b$ excitation pathway.
Within the STD framework, Eq.~\eqref{eq:P_eq} reduces to a polynomial equation for $P_{\mathrm{eq}}$ that depends exclusively on the nuclear spin $I$ (refer to Table~\ref{tab:p_F_S_zF}). 
Notably, higher values of $I$ correspond to a smaller $P_{\mathrm{eq}}$.
This scaling arises because a larger $I$ introduces more $m_F$ sublevels, effectively diluting the initial low-$\Phi$ degeneracy advantage of the $F=a$ multiplet. 
Consequently, the $F=b$ resonance overtakes the $F=a$ pumping efficiency at a lower polarization threshold.

\subsection{\label{sec:magnetic_resonance_enhancement}Magnetic resonance enhancement}

For polarized atoms under RF driving, we investigate the impact of QHP-specific optical pumping on the magnetic-resonance linewidth, $\Gamma_2$, 
which ultimately governs the magnetometer performance discussed in Sec.~\ref{sec:theory_magnetic_resonance}.

Figure~\ref{fig8:linewidth_narrowing}(a) plots $\Gamma_2$ as a function of the incident photon flux $\Phi$ under representative HP and QHP conditions. 
In the zero-light limit ($\Phi \to 0$), the linewidth converges to a baseline dictated entirely by collisional relaxation:
\begin{align}
\Gamma_{2}(\Phi\to 0) &= \frac{1}{q_{\mathrm{SD}}(I)} \Gamma_{\mathrm{SD}} + \frac{1}{q_{\mathrm{EX}}(I)} \Gamma_{\mathrm{EX}},
\label{eq:Gamma2_baseline}
\end{align}
where $q_{\mathrm{SD}}(I)$ and $q_{\mathrm{EX}}(I)$ denote the nuclear-spin-dependent slowing-down factors \cite{Tang2025}:
\begin{align}
q_{\mathrm{SD}}(I) &= \frac{(2I+1)^2}{2 I^2+I+1}, \quad q_{\mathrm{EX}}(I) = \frac{3(2I+1)^2}{2 I(2 I-1)}.
\end{align}
As $\Phi$ increases, optical pumping actively reshapes the resonance linewidth. 
In the HP limit (dashed line), the unresolved hyperfine structure leads to a well-established $\Phi$-dependence: under rapid spin-exchange conditions,
$\Gamma_2$ exhibits initial narrowing followed by severe light-induced broadening at higher polarizations \cite{Tang2025,AppeltNarrowing1999}. 
Specifically, in the strong-pumping limit, this $F$-unresolved HP linewidth asymptotically approaches \cite{SavukovRF2005,Tang2025}
\begin{align}
\Gamma^{\mathrm{(HP)}}_2(P\to 1) = \frac{1}{2I+1}R_{\mathrm{op}} + \frac{1}{2}\Gamma_{\mathrm{SD}}.
\label{eq:Gamma2_strong_pumping}
\end{align}
While the QHP regime (solid curves) qualitatively preserves this initial-narrowing-to-broadening trend, 
it introduces pronounced $F$-resolved characteristics. 
Most notably, across a wide range of $\Phi$, the linewidth obtained by pumping at the $\Delta_b=0$ resonance (blue curve) is consistently narrower than that at $\Delta_a=0$ (red curve). 
This marked disparity indicates that optical pumping via the $F=b$ multiplet inherently provides an additional mechanism for linewidth suppression.

We attribute this pronounced linewidth narrowing to the asymmetric absorption between the $F=a$ and $F=b$ multiplets.
Physically, in a highly polarized atomic ensemble, the Zeeman coherence $\vert \tilde{\rho}_1)$ is predominantly sustained by the atoms residing in the $F=a$ level. 
Consequently, any optical transition event originating from the $F=a$ multiplet directly destroys this coherence, serving as the primary source of light-induced relaxation.
Mathematically, this destructive process rate is scaled by the $F=a$ absorption weight factor, $Q_a$. 
When the pump laser is tuned to the $F=b$ resonance ($\Delta_b = 0$), 
light absorption from the $F=a$ multiplet is heavily off-resonant, driving $Q_a$ to values well below unity (as illustrated in Fig.~\ref{fig2:Q_F}).
Under this condition, the coherent $F=a$ atoms are effectively shielded from optical relaxation. 
Conversely, pumping directly at the $F=a$ resonance ($\Delta_a = 0$) yields $Q_a > 1$, which severely exacerbates decoherence.
This $Q_a$-scaling mechanism becomes glaringly evident in the intense-light limit ($P \to 1$), 
where optical pumping entirely dominates the relaxation process.
In this limit, the magnetic-resonance linewidth asymptotically approaches
\begin{align}
\Gamma_{2}(P\to 1) &= \frac{1}{2I+1} A\Phi\mathcal{L}(\Delta_a) + \frac{1}{2}\Gamma_{\mathrm{SD}},
\label{eq:linewidth_analytic_P0}
\end{align}
a result derived by replacing the $F$-unresolved pumping rate in the HP counterpart Eq.~\eqref{eq:Gamma2_strong_pumping} with the effective $F=a$ absorption rate, $R_{\mathrm{op}}Q_a = A\Phi\mathcal{L}(\Delta_a)$.
This analytic limit perfectly encapsulates the physical intuition: 
operating at the $F=b$ resonance fundamentally decouples the creation of spin polarization from severe destruction of Zeeman coherence, thereby drastically retarding linewidth growth as the photon flux increases.

Originating from the same physical mechanism as the light-induced broadening, 
the pump field modifies the Larmor frequency via the optical shift, $\delta\omega_{\mathrm{L}}$, expressed as
\begin{align}
\delta\omega_{\mathrm{L}} &= \frac{\Delta_a}{\Gamma_{\mathrm{brd}}} \frac{A\Phi\mathcal{L}(\Delta_a)}{2I+1}.
\end{align}
These QHP-specific theoretical predictions regarding both the linewidth $\Gamma_2$ and the optical shift $\delta\omega_{\mathrm{L}}$ consistently explain previous experimental observations \cite{ChangPRA2019,ScholtesMx2011}.
In practical vapor cells, laser intensity gradients render $\delta\omega_{\mathrm{L}}$ spatially dependent, introducing inhomogeneous line broadening that degrades sensor sensitivity. 
While operating exactly at $\Delta_a = 0$ eliminates this shift, it inevitably incurs excessive power broadening as discussed above. 
By contrast, the $F=b$ resonant configuration suppresses the magnitude of $\delta\omega_{\mathrm{L}}$ due to its large hyperfine detuning from the $F=a$ state, thereby minimizing the detrimental impact of spatial gradients.

Furthermore, these mechanistic advantages translate into substantially enhanced magnetic resonance signals $X$ and $Y$,
as illustrated in Figs.~\ref{fig8:linewidth_narrowing}(b) and \ref{fig8:linewidth_narrowing}(c).
Although pumping at the $F=b$ resonance (blue curves) introduces a modest optical shift, it yields a markedly larger resonance amplitude compared to the $F=a$ condition (red curves).
For comparison, the black dashed curves show the HP response at the same photon flux; the resonance is both weaker and broader.
The extent of this enhancement is governed by the equal-polarization threshold, $\Phi_{\mathrm{eq}}$ in Eq.~\eqref{eq:Phi_eq}.
Below this critical flux ($\Phi < \Phi_{\mathrm{eq}}$), the competing dynamics of polarization generation and coherence relaxation result in comparable signal strengths for both conditions.
Conversely, in the strong-pumping regime ($\Phi > \Phi_{\mathrm{eq}}$), the $F=b$ resonance provides a decisive synergistic advantage: 
it sustains a higher macroscopic spin polarization to drive the coherence while simultaneously shielding the system from severe optical broadening, 
thereby optimizing the overall magnetic resonance performance.

Ultimately, the core advantage in the QHP regime lies in its unique optical pumping dynamics: 
resonant excitation at the $F=b$ multiplet actively shields Zeeman coherences from optical relaxation while maintaining robust spin polarization.
This mechanism yields narrower magnetic-resonance linewidths and amplified signal strengths.
Consequently, leveraging these dynamics provides a direct pathway to significantly enhance the intrinsic sensitivity of atomic magnetometers operating at practical buffer-gas pressures. 

\section{\label{sec:conclusions}Conclusions}
We developed a theoretical framework for optical pumping of alkali-metal atoms in buffer-gas cells operating in the QHP regime, where the collisional broadening $\Gamma_{\mathrm{brd}}$ is comparable to the GS hyperfine splitting $\omega^{\{\mathrm{g}\}}_{\mathrm{hf}}$. 
By incorporating the weighting factor $Q_F$ to account for asymmetric absorption, we derived the QHP-specific optical pumping superoperator within the Liouville-space master equation framework.
Its $F$-resolved form established a compact relation linking to the HP counterpart, and facilitates both physical analysis and numerical evaluation in the QHP regime.

Using this unified approach, we demonstrated that QHP optical pumping dynamics are governed by the combined effects of population redistribution across both $F$ multiplets and $m_F$ levels. We presented and discussed these $F$-pumping and $m_F$-pumping effects under different light polarizations. Furthermore, we examined the resulting steady-state populations that deviate from the standard STD and pointed out the validity of the STD approximation under these conditions.

Regarding observables with QHP corrections, we systematically evaluated three interconnected quantities under $F$-resolved pumping: the absorption cross section $\sigma$, the electron spin polarization $P$, and the intrinsic magnetic-resonance linewidth $\Gamma_2$. The characteristic behaviors of absorption and polarization are highlighted by features such as the convergent integral area under linearly polarized pumping, alongside the distinct equal-polarization points under circularly polarized pumping. Applying these insights to atomic magnetometers, we identified an optimal operating scheme, demonstrating that pumping at the $F=b$ resonance comprehensively enhances performance by suppressing the linewidth and amplifying the signal amplitude.

In summary, our study provides a profound understanding of the QHP-specific optical pumping and offers useful theoretical tools with practical applicability to atomic sensors (e.g., atomic magnetometers) operating under realistic buffer-gas pressures. 
The methodology developed herein also establishes a foundation for integrating further physical effects in the QHP regime, including atomic diffusion.

\section*{ACKNOWLEDGMENTS}
This work is supported by Science Challenge Project, No. TZ2025017.

\section*{DATA AVAILABILITY}
The data that support the findings of this article are not
publicly available. The data are available from the authors
upon reasonable request.

\appendix

\section{PROOF OF THE QHP CORRECTION IN LIOUVILLE SPACE}
\label{app:simplification}
Based on the QHP approximation for light detuning [see Eq.~\eqref{eq:QHP_approximation}], we provide a detailed derivation of the $F$-resolved superoperators 
$\mathcal{A}_{\mathrm{dp}}\mathcal{P}_{\mathrm{Z}}$ Eq.~\eqref{eq:A_dp_QHP} 
and $\mathcal{A}^{\{\mathrm{e}g\}}_{\mathrm{p}}\mathcal{P}_{\mathrm{Z}}$ Eq.~\eqref{eq:Ap_eg_QHP}. 
Furthermore, we demonstrate that this compact $F$-resolved formalism does not apply to the hyperfine coherences between the $F=a$ and $F=b$ multiplets.

To simplify the depopulation superoperator, we
utilize the identities in Liouville space \cite{HapperBook2010}, 
\begin{gather}
{X}^{\textcopyright} = {X}^{\flat}- {X^{\dagger\sharp}}, \\
(AB)^{\flat} = A^{\flat}B^{\flat}, \quad  (AB)^{\sharp} = B^{\sharp}A^{\sharp}, 
\end{gather}
where $A$, $B$, and $X$ are Hilbert-space operators. Consequently, by substituting the $F$-resolved $\delta H^{\{\mathrm{g}\}}$ from Eq.~\eqref{eq:delta_H_QHP} into Eq.~\eqref{eq:dp_def}, the superoperator $\mathcal{A}_{\mathrm{dp}}$ becomes 
\begin{align}
\mathcal{A}_{\mathrm{dp}}&= \frac{i}{R_{\mathrm{op}}}\sum_{F} \left(\frac{\left(V^{\dagger}V \right) ^{\flat}\Pi_F^{\flat}}{\Delta_F + i\Gamma_{\mathrm{brd}}}  - \frac{\left( V^{\dagger}V  \right)^{\sharp}\Pi_F^{\sharp}}{\Delta_F - i\Gamma_{\mathrm{brd}}}\right).
\end{align}
The projection superoperators in Liouville space are related to those in Hilbert space via
\begin{align}
\Pi_{F}^{\flat} &= \mathcal{P}_{F}+ \mathcal{P}_{F\to F'},\\ 
\Pi_{F}^{\sharp} &= \mathcal{P}_{F}+ \mathcal{P}_{F'\to F}.
\end{align}
Here, the projection superoperators for the hyperfine coherences ($F\neq F'$) are defined as
\begin{align}
\mathcal{P}_{F\to F'} &= \sum_{m_{F},m'_{F}}\left|F_{},m_{F_{}};F'_{},m'_{F_{}}\right)\left(F_{},m_{F_{}};F'_{},m'_{F_{}}\right|,\\
\mathcal{P}_{\mathrm{hf}}&=\mathcal{P}_{a\to b}+\mathcal{P}_{b\to a}.
\end{align}
Applying these projection superoperator properties, we obtain 
\begin{align}
\mathcal{A}_{\mathrm{dp}}\mathcal{P}_{\mathrm{Z}}
&= \frac{i}{R_{\mathrm{op}}}\sum_{F} \left( \frac{\left(V^{\dagger}V \right)  ^{\flat}}{\Delta_F + i\Gamma_{\mathrm{brd}}}  - \frac{\left(V^{\dagger}V \right) ^{\sharp}}{\Delta_F - i\Gamma_{\mathrm{brd}}} \right)\mathcal{P}_{F},\label{eq:A_dp_QHP_app} \\
\mathcal{A}_{\mathrm{dp}}\mathcal{P}_{\mathrm{hf}}
&= \frac{i}{R_{\mathrm{op}}}\sum_{F} \left(\frac{\left(V^{\dagger}V \right) ^{\flat}\mathcal{P}_{F\to F'}}{\Delta_F + i\Gamma_{\mathrm{brd}}}  - \frac{\left( V^{\dagger}V  \right)^{\sharp}\mathcal{P}_{F'\to F}}{\Delta_F - i\Gamma_{\mathrm{brd}}}\right). \label{eq:A_dp_hf}
\end{align}
Equation~\eqref{eq:A_dp_QHP_app} demonstrates that, under the QHP correction, $\mathcal{A}_{\mathrm{dp}}\mathcal{P}_{\mathrm{Z}}$ acquires a distinct $F$-resolved structure. This allows the depopulation dynamics to be decoupled into separate contributions from the $F=a$ and $F=b$ manifolds, mathematically analogous to the behavior of $\delta H^{\{\mathrm{g}\}}$ in Eq.~\eqref{eq:delta_H_QHP}. Conversely, Eq.~\eqref{eq:A_dp_hf} reveals that the respective contributions from the positive- and negative-frequency hyperfine coherences to the depopulation dynamics cannot be unified into such a compact form.

To further simplify the expression, we relate the electric-dipole interaction strength to the photon flux via $|\Omega_{\mathrm{I}}|^{2}=A\Phi$. This naturally introduces the dimensionless weight factor:
\begin{align}
Q_{F}\equiv \frac{|\Omega_{\mathrm{I}}|^{2}\mathcal{L}(\Delta_F)}{R_{\mathrm{op}}} =\frac{\mathcal{L}(\Delta_F)}{x_a\mathcal{L}(\Delta_a) + x_b\mathcal{L}(\Delta_b)}. 
\label{eq:Q_F_Omega_I}
\end{align}
By utilizing the symmetry property of the dipole operator, $\hat{\mathbf{e}}^* \cdot \mathbf{d} \mathbf{d}^{\dagger} \cdot \hat{\mathbf{e}} = 1/2 - K\mathbf{s} \cdot \mathbf{S}$, we arrive at the compact form of Eq.~\eqref{eq:A_dp_QHP_app}:
\begin{align}
\mathcal{A}_{\mathrm{dp}}\mathcal{P}_{\mathrm{Z}}
&= \sum_{F} \left(\mathrm{Re}\Big[\mathcal{A}^{(\mathrm{HP})}_{\mathrm{dp}}\Big] {Q}_{F} 
+i\mathrm{Im}\Big[\mathcal{A}^{(\mathrm{HP})}_{\mathrm{dp}}\Big] \frac{{\Delta_{F}Q_{F}}}{\Delta_0}\right) \mathcal{P}_F.
\end{align}
This result completes the proof of Eq.~\eqref{eq:A_dp_QHP}. 
Structurally, this QHP-corrected expression builds directly upon the traditional HP superoperator $\mathcal{A}^{(\mathrm{HP})}_{\mathrm{dp}}$ [Eq.~\eqref{eq:A_dp_HP}] by incorporating the projection superoperator $\mathcal{P}_F$ and the weight factor $Q_F$, which exactly accounts for the asymmetric light absorption between different hyperfine levels.

We now turn to the optical excitation superoperator $\mathcal{A}^{\{\mathrm{eg}\}}_{\mathrm{p}}$, focusing on deriving its $F$-resolved form. Following a similar procedure and starting from its definition in Eq.~\eqref{eq:Ap_eg_def}, we obtain
\begin{align}
\mathcal{A}^{\{\mathrm{eg}\}}_{\mathrm{p}} \mathcal{P}_{\mathrm{Z}}
&= \frac{i}{R_{\mathrm{op}}}\sum_{F} \left( \frac{V^{\dagger\sharp}V^{\flat}}{\Delta_F + i\Gamma_{\mathrm{brd}}}  -  \frac{V^{\dagger\sharp}V ^{\flat}}{\Delta_F - i\Gamma_{\mathrm{brd}}} \right)\mathcal{P}_{F}.
\label{eq:Ap_eg_QHP_app}
\end{align}
Similar to Eq.~\eqref{eq:A_dp_hf}, this $F$-resolved decomposition is also invalid for hyperfine coherences. 
By applying the Liouville-space property ${B}^{\mathrm{T}} \otimes A=A^{\flat}B^{\sharp}$ \cite{HapperBook2010}, the superoperator can be recast into the final QHP-corrected form:
\begin{align}
\mathcal{A}^{\{\mathrm{eg}\}}_{\mathrm{p}}\mathcal{P}_{\mathrm{Z}}
&= \sum_{F}  A^{\{\mathrm{eg}\}(\mathrm{HP})}_{\mathrm{p}}
{Q}_{F} \mathcal{P}_F.
\end{align}
Here, the HP counterpart $\mathcal{A}^{\{\mathrm{eg}\}(\mathrm{HP})}_{\mathrm{p}}$ governs the $F$-unresolved optical excitation dynamics, recovering the classical HP limit when $Q_F \to 1$:
\begin{align}
\mathcal{A}^{\{\mathrm{eg}\}(\mathrm{HP})}_{\mathrm{p}} &=  2(\hat{\mathbf{e}}^{*}\cdot \mathbf{d})\otimes (\mathbf{d}^{\dagger}\cdot\mathbf{\hat{e}}).
\end{align}

In summary, the QHP corrections derived for the individual superoperators consistently adhere to a unified $F$-resolved formalism weighted by $Q_F$. This structure not only establishes a rigorous connection to the traditional HP limit but is also mathematically essential for constructing a compact expression for the overall optical pumping superoperator $\mathcal{A}_{\mathrm{op}}$.

\section{FREQUENCY BLOCKS IN ZEEMAN SUBSPACE}
\label{app:blocks}
In Liouville-space calculations for the master equation, an appropriate ordering of the basis states is essential for exposing the block structure of large superoperators. Such a structure not only improves computational efficiency but also highlights the underlying physical organization of the dynamics. A block group scheme is constructed in Zeeman subspace, i.e., $\mathrm{span}\{|Fm_F,Fm'_F)\}$, according to the characteristic evolution frequencies of the density-matrix elements \cite{Tang2025}. 

The projection superoperators of the GS $k$th-order Zeeman coherence are introduced as
\begin{align}
\label{eq:proj_Zeeman}
\mathcal{P}^{\{k\}}_{\mathrm{Z}} &=  \mathcal{P}^{\{k\}}_{a}+\mathcal{P}^{\{k\}}_{b}, \\
\label{eq:proj_a}
\mathcal{P}^{\{k\}}_{a}&=\sum_{m_{F}} \left|a,{m}_{F}+k;a,{m}_{F}\right)\left(a,{m}_{F}+k;a,{m}_{F}\right|, \\
\label{eq:proj_b}
\mathcal{P}^{\{k\}}_{b}&=\sum_{m_{F}}\left|b, {m}_{F}-k;b,{m}_{F}\right)\left(b,{m}_{F}-k;b,{m}_{F}\right|.
\end{align}
where $\left|F,m_{F};F',m'_{F}\right)$ is the basis of $|{\rho})$ in Liouville space,
\begin{align}
\label{eq:basis_def}
& |F,m_F;F',m'_{F}) = |F, m_F\rangle\langle F', m'_{F}|.
\end{align}
Using the projection superoperators, the vectorized density matrix $|{\rho})$ in a weak magnetic field $B$ can be grouped into $k$th-order Zeeman components $|{\rho}^{\{k\}}_{\mathrm{Z}})=\mathcal{P}^{\{k\}}_{\mathrm{Z}}\big|{\rho}\big)$.  
Naturally, the zeroth-order Zeeman coherence is population $|\rho_0)=|{\rho}^{\{0\}}_{\mathrm{Z}})$. 
For $k\neq 0$, it is physically imperative to note that the projection definitions for the $F=a$ and $F=b$ manifolds shift the magnetic quantum number in opposite directions ($+k$ versus $-k$). This deliberate construction compensates for their oppositely signed gyromagnetic ratios and Zeeman shifts. Consequently, all elements within a given $|{\rho}^{\{k\}}_{\mathrm{Z}})$ subspace evolve at a common precession frequency $\vert \omega_{{F},{m}_F+k}^{\{\mathrm{g}\}} - \omega_{F, m_F}^{\{\mathrm{g}\}}\vert$, significantly simplifying dynamical evolution.

Following this frequency-matched scheme, for any general superoperator $\mathcal{X}$ acting on $|{\rho})$, we partition it into corresponding block matrices
\begin{align}
\mathcal{X}&=\sum_{m,n}[\mathcal{X}]_{m, n},\\
[\mathcal{X}]_{m, n}&=\mathcal{P}^{\{m\}}_{\mathrm{Z}}[\mathcal{X}]\mathcal{P}^{\{n\}}_{\mathrm{Z}},
\end{align}
where $[\mathcal{X}]_{m, n}$ physically describes the dynamical coupling from the $n$th-order to the $m$th-order Zeeman coherence subspace.

To further streamline the full dynamical evolution of $\left|{\rho}_{}\right)$,
we neglect the coherences between different hyperfine multiplets and restrict our focus to mechanisms within the Zeeman subspace.
This approximation is highly justified because, in the absence of external microwave driving, 
hyperfine coherences rapidly decay 
on a timescale typically on the order of milliseconds or shorter \cite{CorsiniTimescale2013}.

\section{\label{app:D2_pumping}OPTICAL PUMPING OF THE D$_2$ LINE}

This appendix addresses the dynamics of macroscopic observables under the D$_2$-line optical pumping in the QHP regime. 
Fundamentally, pumping on the D$_1$ ($K=1$) and D$_2$ ($K=-1/2$) transitions engages distinct GS-ES coupling pathways, dictating different underlying mechanisms \cite{HapperRMP1972,Seltzer2008}. 
Specifically, while the D$_1$ pumping employs the formation of dark states to achieve highly efficient spin polarization, the D$_2$ pumping relies on differential absorption rates among $m_F$ sublevels, yielding more intricate dynamics and inherently lower pumping efficiency. 
Although the foundational principles of the QHP corrections apply universally to both transitions, the specific steady state of atomic spin differs substantially. To elucidate D$_2$-specific characteristics, we detail its distinct behaviors regarding light absorption and spin polarization.

\begin{figure}[b]
\includegraphics[width=1\linewidth]{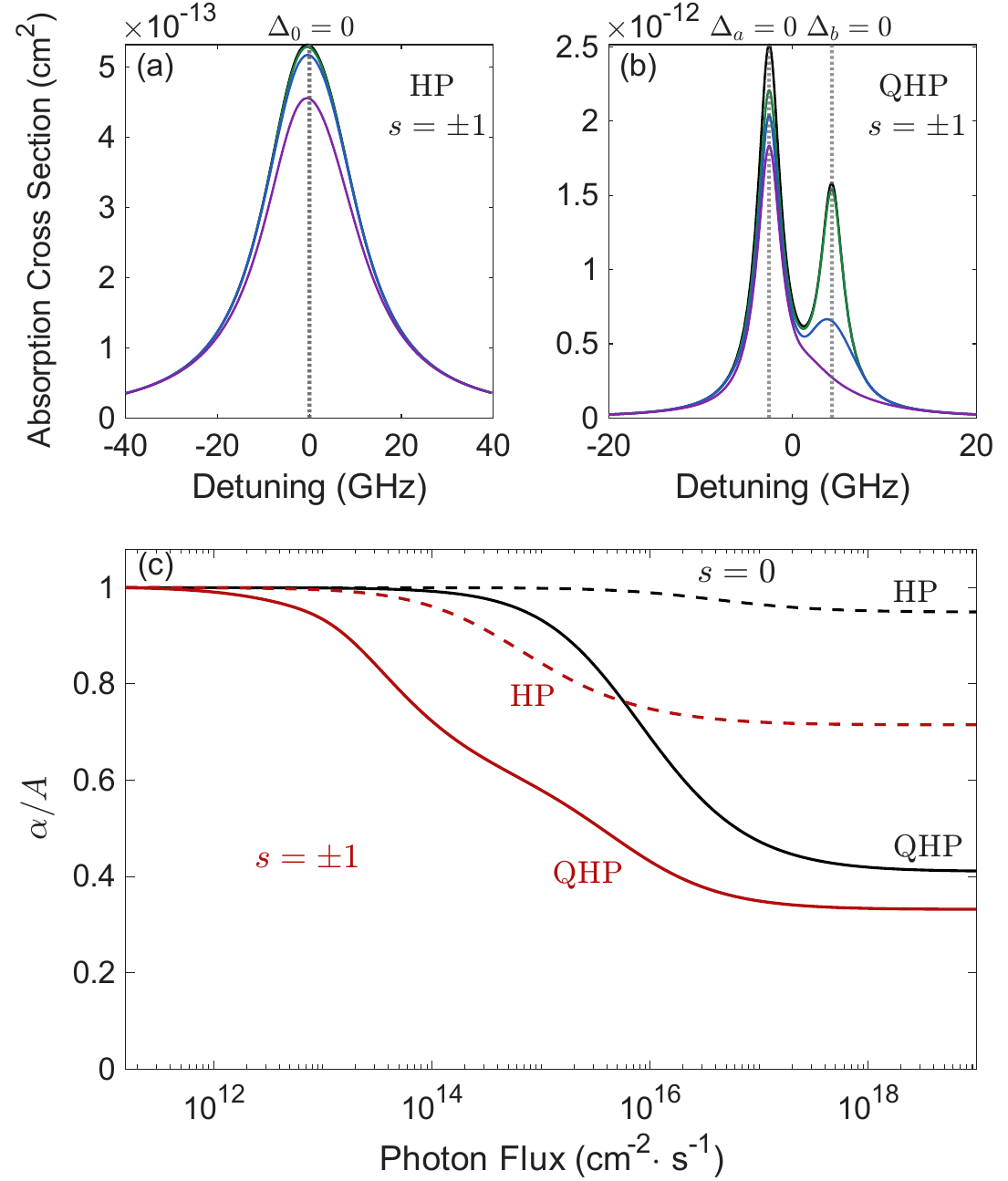} 
\caption{
Absorption cross section of the D$_2$ line under circularly polarized pumping. (a) and (b) represent the HP and QHP regimes, respectively. Color coding and calculation parameters match Figs.~\ref{fig5:absorption_cross_section}(c) and \ref{fig5:absorption_cross_section}(d) for direct comparison. (c) displays the corresponding integrated area $\alpha$ of the D$_2$ line. Note that the linearly polarized pumping case ($s=0$) is identical to that in Fig.~\ref{fig6:integrated_sigma}.
}
\label{fig9:D2_absorption_cross_section} 
\end{figure}

Figure~\ref{fig9:D2_absorption_cross_section} illustrates the D$_2$ line absorption cross section in both HP and QHP regimes. We highlight the distinct behaviors under circularly polarized pumping, with the linearly polarized pumping case identical to that in Sec.~\ref{sec:light_absorption_cross_section}. 
Compared with the D$_1$ pumping, in the HP regime, Fig.~\ref{fig9:D2_absorption_cross_section}(a) exhibits a weaker suppression of overall absorption as pumping intensity increases.
This feature becomes $F$-resolved in the QHP regime [Fig.~\ref{fig9:D2_absorption_cross_section}(b)]: although absorption at the $\Delta_b=0$ resonance weakens, the $F=a$ resonance maintains a distinct and persistent absorption peak.
Consequently, as shown in Fig.~\ref{fig9:D2_absorption_cross_section}(c), the integrated spectral area $\alpha$ never fully vanishes under intense pumping in either regime (red curves), although it converges to a lower finite value under QHP conditions.

These observed differences originate from the fundamental absence of dark states in D$_2$ transitions, compelling all atomic states---particularly within the $F=a$ multiplet---to maintain residual light absorption.
This renders the optical pumping less effective at globally suppressing absorption.
Specifically, in the HP regime, the presence of these bright sublevels yields $\sigma_{\mathrm{cir}} = \bar{\sigma}_{\mathrm{HP}}(1 - |P_{\mathrm{HP}}|/2)$; since the maximum achievable polarization is limited to $|P_{\mathrm{HP}}|=1/2$, the integrated area naturally converges to $3A/4$ (red dashed curve). 
In the QHP regime, the inherently bright $F=a$ multiplet consistently contributes to a significant absorption background.
Thus, a distinct decrease in absorption occurs exclusively under resonant $F=b$ pumping.

Another noteworthy feature for the D$_2$ line is that the QHP-specific $\alpha$ exhibits a multi-stage behavior [red solid curve], where the turning domain signifies the invalidity of the STD approximation. 
Mechanistically, this arises because the $m_F$-pumping and $F$-pumping dominate in different intensity regimes: while $m_F$-pumping governs the initial stage, strong $F$-pumping with all levels bright overwhelms the spin-exchange rate at higher $\Phi$ and breaks down the STD approximation. 
Consequently, this transition precludes a general analytical expression for the cross section.

\begin{figure}
\includegraphics[width=1\linewidth]{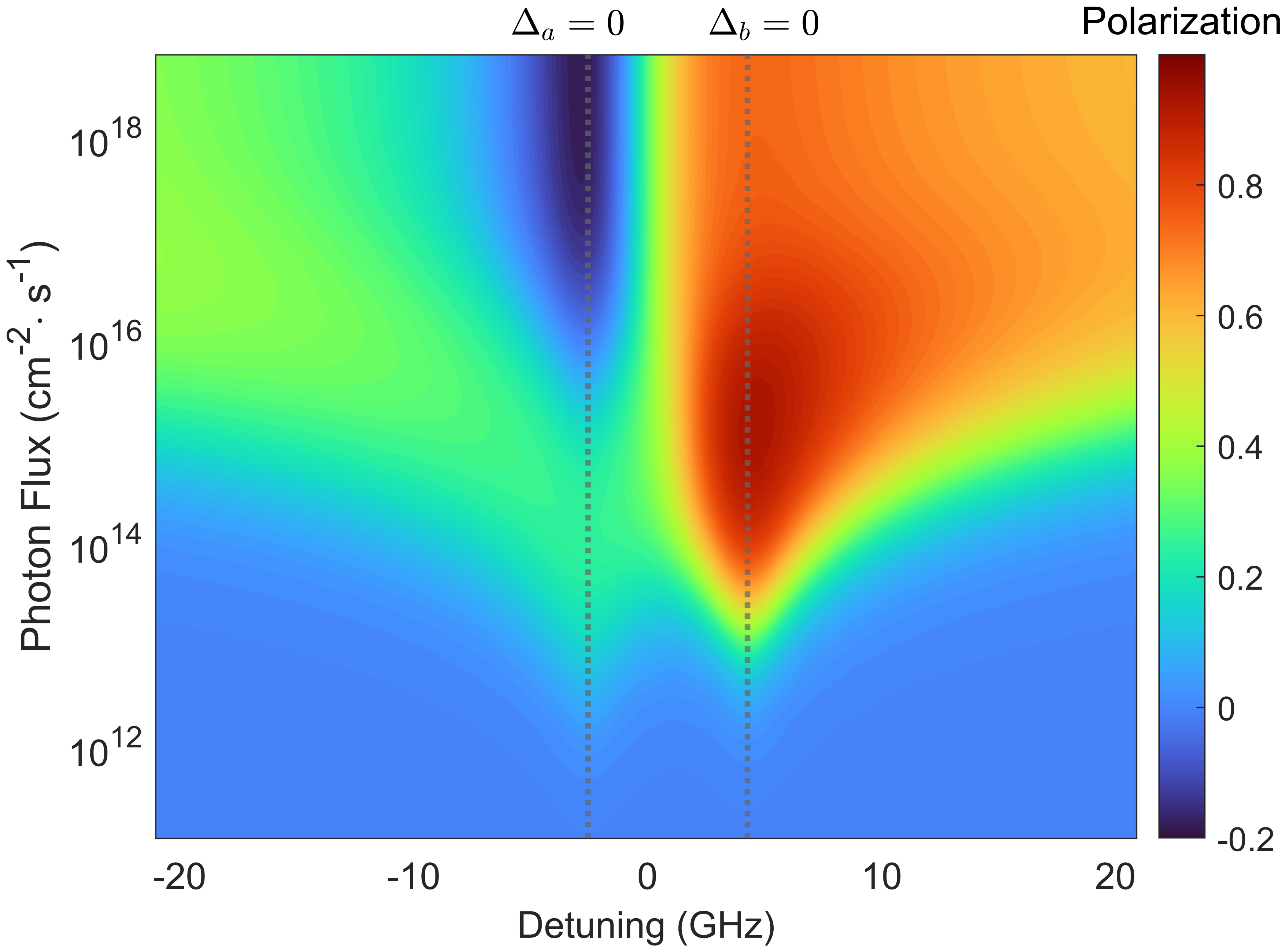} 
\caption{
Steady-state spin polarization $P$ under D$_2$-line pumping with $s=-1$. The collisional broadening and relaxation rates are set identical to those in Fig.~\ref{fig7:polarization_sweeper}.
}
\label{fig10:D2_polarization_sweeper} 
\end{figure}

The invalidity of the STD approximation under strong pumping is directly manifested in the spin polarization $P$ depicted in Fig.~\ref{fig10:D2_polarization_sweeper}.
When tuning the pumping to the $F=a$ resonance, $P$ behaves non-monotonically with increasing $\Phi$: it initially builds up, subsequently degrades, and ultimately undergoes a significant reversal.
This anomalous result occurs because strong $F$-pumping predominantly populates the $F=b$ multiplet and forces atoms to accumulate in the sublevels ranging from $m_F = -s$ to $-sb$, which leads to a reversal of electron spin polarization. 
When transitioning to the $F=b$ resonance, the population in the stretched state is better preserved, allowing for further optimization of pumping efficiency.

Despite these mechanistic differences in absorption and polarization between the D$_1$ and D$_2$ lines, the advantageous linewidth narrowing and signal enhancement for the $F=b$ resonance discussed in Sec.~\ref{sec:magnetic_resonance_enhancement} remain applicable.
Even with the complex dynamics and inherently lower pumping efficiency of the D$_2$ transition, this specific configuration still successfully circumvents the dominant coherence relaxation associated with the $F=a$ multiplet while optimizing the accessible degree of spin polarization.

\makeatletter
\newwrite\mycontrolbib
\immediate \openout \mycontrolbib = \jobname Notes.bib\relax
\immediate\write\mycontrolbib{@CONTROL{apsrev42Control, title="0"}}
\immediate\closeout\mycontrolbib
\immediate\write\@auxout{\string\citation{apsrev42Control}}
\makeatother

\bibliographystyle{apsrev4-2}
\bibliography{references}

\end{document}